# Networks as Proxies: a relational approach towards economic complexity in the Roman period*

Johannes Preiser-Kapeller, Institute for Medieval Research, Division of Byzantine Research, Austrian Academy of Sciences
Email: Johannes.Preiser-Kapeller@oeaw.ac.at
Website: http://oeaw.academia.edu/JohannesPreiserKapeller

**Abstract:** Based on the assumption that economic complexity is characterised by the interactions of "economic agents (who) constantly change their actions and strategies in response to the outcome they mutually create", this paper presents how network models can be used a "proxies" for the mapping, quantification and analysis of Roman economic complexity. Network analysis provides tools to visualise and analyse the "inherent" complexity of various types of data and their combination (archaeological, geographical, textual) or even of a single piece of evidence. Equally, the relational approach invites to a structural and quantitative comparison between periods, regions and the economic systems of polities and empires. An increasing number of proxies of this kind may allow us to capture the trajectories of economic complexity beyond metaphors.

## 1 Systems, complexity and networks

"Complexity economics" has become a prominent topic over the last three decades; as W. Brian Arthur, one of the pioneers of this field, explains: "complexity economic builds on the proposition that the economy is not necessarily in equilibrium: economic agents (firms, consumers, investors) constantly change their actions and strategies in response to the outcome they mutually create. (…) Complexity economic thus sees the economy in motion, perpetually "computing" itself – perpetually constructing itself anew. Where equilibrium economics emphasizes order, determinacy, deduction, and stasis, complexity economics emphasizes contingency, indeterminacy, sense-making, and openness to change."[1]

As Arthur is referring primarily to modern economic dynamics, one could ask how much place there was for "contingency, indeterminacy, sense-making, and openness to change" in pre-modern economic frameworks such as the Roman one. In any case, notions of "system"[2]

---

* This paper was prepared within the project "Harbours and landing places on the Balkan coasts of the Byzantine Empire (4th to 12th centuries)" (part of the SPP-1630 "Harbours from the Roman period to the Middle Ages", funded by the Deutsche Forschungsgemeinschaft); the project is undertaken at the Römisch-Germanisches Zentralmuseum (RGZM) in Mainz in cooperation with the Institute for Byzantine and Modern Greek Studies of the University of Vienna and the Institute for Medieval Research, Division of Byzantine Research, Austrian Academy of Sciences. Project leader is Prof. Falko Daim (RGZM).

[1] W. B. Arthur, Complexity and the Economy. Oxford 2015, 1, see also 89-102 on "process and emergence in economy". Cf. also B. C. Beaudreau, On the Emergence and Evolution of Economic Complexity. *Modern Economy* 2 (2011) 266-278.

[2] For the notion of "system" with regard to economy, its "autopoiesis" and its relations to other "social systems" cf. N. Luhmann, Die Wirtschaft der Gesellschaft. Frankfurt am Main 1994, esp. 43-90: Luhmann cites H. R. Maturana (Autopoiesis, in: M. Zeleney [ed.], Autopoiesis: a theory of living organization. New York 1981, 21)

and "complexity"[3] as well as "network" (whose study "for many scientists in the community (…) is synonymous with the study of complexity"[4]) are very much present in historical and even more in archaeological papers for some time now. While in many studies, these terms are used in a more "metaphorical" way or as novel conceptual framework for otherwise traditional narratives[5], a considerable number of scholars actually refer also to their formal and mathematical basis.

Among these approaches we find:

- Attempts to identify statistical "signatures of complexity" in quantitative data (e. g., distributions of settlements sizes within a region or of wealth among individuals in a larger group, respectively time series, esp. of prices, but also of proxy data), such as unequal distributions patterns (power law, logarithmic, etc.) or indicators for non-linear dynamics (Lyapunov-exponent, etc.).[6] Furthermore, systems of equations are proposed in order to capture essential factors for these dynamics on the basis of the

---

for a definition of autopoiesis: "there are systems that are defined as unities of networks of productions of components that (1) recursively, through their interactions, generate and realize the network that produces them; and (2) constitute, in the space in which they exist, the boundaries of this network as components that participate in the realization of the network."

[3] As Arthur (Complexity and the Economy 3) highlights, "complexity is not a theory but a movement in the sciences that studies how the interacting elements in a system create overall patterns, and how these overall patterns in turn cause the interacting elements to change or adapt". Cf. also Kl. Mainzer, Thinking in Complexity. The Computational Dynamics of Matter, Mind, and Mankind. 5th ed., Berlin – Heidelberg 2007; R. A. Bentley – H. D. G. Maschner, Complex systems and archaeology. Salt Lake City 2003; T. A. Kohler, Complex systems and archaeology, in: I. Hodder (ed.), Archaeological theory today. Cambridge 2012, 93-123. For a good introduction for historians see J. L. Gaddis, The Landscape of History. How Historians map the Past. Oxford 2002.

[4] N. Johnson, Simply Complexity. A clear Guide to Complexity Theory. London 2009, 13. For an overview of network analytical approaches to economic complexity cf. A. Chaterjee – B. K. Chakrabarti (eds.), Econophysics of Markets and Business Networks. Heidelberg – New York 2007; M. O. Jackson, Social and Economic Networks. Princeton – Oxford 2008; S. Sinha et al., Econophysics. Weinheim 2010, esp. 203-243; D. Easly – J. Kleinberg, Networks, Crowds, and Markets. Reasoning about a Highly Connected World. Cambridge 2010; D. Knoke, Economic Networks. Cambridge 2012.

[5] Cf. for instance I. Malkin, A Small Greek World. Networks in the Ancient Mediterranean. Oxford 2011, and most contributions in: I. Malkin – Ch. Constantakopoulou – K. Panagopoulou (eds.), Greek and Roman Networks in the Mediterranean. London – New York 2009.

[6] Cf. Th. A. Brown, Measuring Chaos using the Laypunov-Exponent, in: L. D. Kiel – E. Elliott (eds.), Chaos Theory in the Social Sciences. Foundations and Applications. Ann Arbor 1997, 53-66; H. Kantz – Th. Schreiber, Nonlinear Time Series Analysis. Cambridge ²2004; H. Thome, Zeitreihenanalyse. Eine Einführung für Sozialwissenschaftler und Historiker. Munich – Vienna 2005; Sinha et al., Econophysics 83-129 (on distribution patterns). For a survey on studies producing proxy data for economic growth in ancient Greece cf. for instance J. Ober, Wealthy Hellas. *Journal of Economic Asymmetries* 8/1 (June 2011) 1-38, and for Rome: A. Wilson, Quantifying Roman economic performance by means of proxies: pitfalls and potential, in: F. de Callataÿ (ed.), Long-Term Quantification in Ancient Mediterranean History (*Pragmateiai* 27). Bari 2014, 147-167; A. Bowman – A. Wilson, Quantifying the Roman Economy: Integration, Growth, Decline?, in: A. Bowman – A. Wilson (eds.), Quantifying the Roman Economy: Methods and Problems. Oxford 2009, 3-83; W. M. Jongman, Reconstructing the Roman economy, in: L. Neal – J. G. Williamson, The Cambridge History of Capitalism. Volume 1. The Rise of Capitalism: From Ancient Origins to 1848. Cambridge 2014, 75-100.

correspondence between patterns emerging from these models and observed data (a top-down approach).[7]

- Efforts to survey, map and analyse the connections and interactions between various elements (individuals, groups, settlements, polities, but also objects or semantic entities) documented in historical or archaeological evidence with the help of network models in the form of graphs with "nodes" and "ties", also in their spatial and temporal dynamics. Again, statistical "signatures of complexity" (e. g. patterns of distribution of the number of links among nodes) are identified and models for their emergence in growing or changing networks are proposed (e. g., mechanisms of preferential attachment causing increasing inequality among nodes regarding their "centrality") (see below for a further outline).

- Experiments to capture the "bottom up"-dynamics of complex systems emerging from the interaction of single elements with the help of agent-based models, acting on the basis of a set of (often relatively simple) rules within a simulated (spatial) environment over several time steps.[8] Again, emerging statistical properties of such models are compared with observed data in order to determine their explanatory value.[9]

In several cases, these approaches are combined.[10] Yet despite all mathematical and computer-based sophistication, similar to any other study on the past they depend on the density and quality of historical (mainly textual) or archaeological evidence. And while they may provide insights into processes and patterns otherwise "in-visible" for any conventional analysis, any evaluation of their heuristic explanatory value relies on their check against "real" data. As I will demonstrate, network analysis provides tools to visualise and analyse the "inherent" complexity of various types of "real" historical and archaeological data and their combination (archaeological, geographical, textual) or even of a single piece of evidence, thus producing a wide range of "proxies for complexity".

---

[7] Cf. W. Weidlich, Sociodynamics: A Systemic Approach to Mathematical Modelling in the Social Sciences. Mineola, New York 2006. For application on historical phenomena see also P. Turchin, Historical Dynamics: Why States Rise and Fall. Princeton 2003.
[8] For the actual "complexity" of economic behaviour of individual agents cf. N. Wilkinson, An introduction to behavioural economics. New York 2008. For the ancient economy cf. B. Schefold, The Applicability of Modern Economics to Forms of Capitalism in Antiquity: Some Theoretical Considerations and Textual Evidence. *The Journal of Economic Asymmetries* 8/1 (June 2011) 131–163, who argues "that there are still good reasons to regard the economic rationality of the ancients as sufficiently different from ours to expect differences between mainstream economics and analytical or verbal approximations to the economics of antiquity in the description of economic processes."
[9] Cf. now M. Madella – B. Rondelli, Simulating the Past: Exploring Change Through Computer Simulation in Archaeology. A special issue of the *Journal of Archaeological Method and Theory* 21 (2) (2014). For a more general overview cf. Sinha et al., Econophysics 147-203 (with further literature).
[10] See for instance several of the contributions in: A. Collar – F. Coward – T. Brughmans – B. J. Mills, The Connected Past: critical and innovative approaches to networks in archaeology. A special issue of the *Journal of Archaeological Method and Theory* 22 (1) (2015).

**1.1 The relational approach: how to model and analyse networks and to measure their complexity**

In general, network theory assumes "not only that ties matter, but that they are organized in a significant way, that this or that (node) has an interesting position in terms of its ties."[11] One central aim of network analysis is the identification of structures of relations which emerge from the sum of interactions and connections between individual, groups or sites and at the same time influence the scope of actions of everyone entangled in such relations. For this purpose, data on the categories, intensity, frequency and dynamics of interactions and relations between entities of interest is systematically collected in a way which allows for further mathematical analysis. This information is organised in the form of matrices (with rows and columns) and graphs (with nodes [vertices] and edges [links]), which are not only instruments of data collection and visualisation, but also the basis of further mathematical operations (on the basis of matrix algebra and graph theory).[12]

Once a quantifiable network model has been created, it allows for a structural analysis on three main levels[13]:

\* the level of single nodes; respective measures take into account the immediate "neighbourhood" of a node – such as "degree", which measures the number of direct links of a node to other nodes[14] or the relative centrality of a node within the entire network due to its

---

[11] Cl. Lemercier, Formale Methoden der Netzwerkanalyse in den Geschichtswissenschaften: Warum und Wie?, in: A. Müller – W. Neurath (eds.), Historische Netzwerkanalysen, Österreichische Zeitschrift für Geschichtswissenschaften 23/1 (Innsbruck, Vienna, Bozen 2012) 16–41, here 22. Cf. also T. Brughmans, Thinking through networks: a review of formal network methods in archaeology. *Journal of Archaeological Method and Theory* 20 (2012) 623-662, the contributions in Ch. Knappett (ed.), Network-Analysis in Archaeology. New Approaches to Regional Interaction. Oxford 2013, and A. Collar – F. Coward – T. Brughmans – B. J. Mills, Networks in Archaeology: Phenomena, Abstraction, Representation, in: A. Collar – F. Coward – T. Brughmans – B. J. Mills, The Connected Past: critical and innovative approaches to networks in archaeology. A special issue of the *Journal of Archaeological Method and Theory* 22 (1) (2015) 1-31, for an overview of concepts and tools as well as further bibliography, as well as T. Brughmans – A. Collar – F. Coward (eds.), The Connected Past. Challenges to Network Studies in Archaeology and History. Oxford 2016. For basic ideas of network theory see also Knoke, Economic Networks 21-24.
[12] St. Wassermann – K. Faust, Social Network Analysis: Methods and Applications, Structural Analysis in the Social Sciences, Cambridge 1994, 92-166; Ch. Prell, Social Network Analysis. History, Theory and Methodology. Los Angeles – London 2012, 9-16; A.-L. Barabási, Network Science. Cambridge 2016, 42-67; J. A. Fuhse, Soziale Netzwerke. Konzepte und Forschungsmethoden. Konstanz – Munich 2016, 41-57.
[13] Collar – Coward – Brughmans – Mills, Networks in Archaeology 17-25, includes also a most useful glossary of basic terms and concepts of network analysis as well as a well-balanced discussion of potential and pitfalls of network models in archaeology. For historical studies the best discussion in this regard is Lemercier, Formale Methoden der Netzwerkanalyse.
[14] Wassermann – Faust, Social Network Analysis 178-183; W. de Nooy – A. Mrvar –Vl. Batagelj, Exploratory Social Network Analysis with Pajek (Structural Analysis in the Social Sciences). Cambridge 2005, 63-64; M.

position on many or few possible paths between nodes otherwise unconnected – the measure of "betweenness", which can be interpreted as a potential for intermediation.[15] A further important indicator of centrality is "closeness", which measures the length of all paths between a node and all other nodes. The "closer" a node is, the lower is its total and average distance to all other nodes. Closeness can also be used as a measure of how fast it would take to spread resources or information from a node to all other nodes.[16]

* the level of groups of nodes, especially the identification of "clusters", meaning the existence of groups of nodes more densely connected to each other than to the rest of the network; if all nodes within such a group are directly connected with each other, they are called "clique"; a measure of the degree to which nodes in a graph tend to cluster together is the "clustering coefficient" (with values between 0 and 1).[17] In order to detect such cliques and clusters, an inspection of a visualisation of a network can be already quite helpful; common visualisation tools arrange nodes more closely connected near to each other ("spring embedder"-algorithms) and thus provide a good impression of such structures.[18] For exact identification, there exist various algorithms of "group detection" (such as the ones developed by the physicist M. Newman, see below), which aim at an optimal "partition" of the network. It is of course also of interest to see if the presence of nodes within such clusters can be related to specific qualitative attributes, for instance.[19] A different approach is the concept of "structural equivalence" of nodes; here, nodes are not attributed to the same "block" because of being connected to each other, but due to having the same (or very similar) structure of ties to other actors (thus, within a network of a school, one would encounter a block of "teachers" and one of "disciples", between which similar structures of relations could be identified). Again, several tools of "blockmodelling" exist.[20]

---

Newman, Networks. An Introduction. Oxford 2010, 168-169; Prell, Social Network Analysis 96-99; Sinha et al., Econophysics 208.

[15] R. S. Burt, Brokerage and Closure: An Introduction to Social Capital. Oxford 2005; Wassermann – Faust, Social Network Analysis 188-192; de Nooy – Mrvar – Batagelj, Exploratory Social Network Analysis 131-133; Newman, Networks 185-193; Prell, Social Network Analysis 103-107.

[16] Wassermann – Faust, Social Network Analysis 184-188; Prell, Social Network Analysis 107-109.

[17] Wassermann – Faust, Social Network Analysis 254-257; Fuhse, Soziale Netzwerke 74-78.

[18] Cf. L. Krempel, Visualisierung komplexer Strukturen. Grundlagen der Darstellung mehrdimensionaler Netzwerke. Frankfurt – New York 2005; D. Dorling, The Visualization of Spatial Social Structure. Chichester 2012.

[19] de Nooy – Mrvar – Batagelj, Exploratory Social Network Analysis 66-77; Newman, Networks 372-382; Prell, Social Network Analysis 151-161; Ch. Kadushin, Understanding Social Networks. Theories, Concepts, and Findings. Oxford 2012, 46-49.

[20] Wassermann – Faust, Social Network Analysis 461-493; de Nooy – Mrvar – Batagelj, Exploratory Social Network Analysis 259-285; Prell, Social Network Analysis 176-194. For an application on historical data cf. J. F. Padgett - Ch. K. Ansell, Robust Action and the Rise of the Medici, 1400–1434. *The American Journal of Sociology* 98/6 (1993) 1259-1319.

\* the level of the entire network: basic key figures are the number of nodes and of links, the maximum distance between two nodes (expressed in the number of links necessary to find a path from one to the other; "diameter") and the average distance (or path length) between two nodes. A low average path length among nodes together with a high clustering coefficient can be connected to the model of a "small world network", in which most nodes are linked to each other via a relatively small number of edges.[21] "Density" indicates the ratio of possible links actually present in a network: theoretically, all nodes in a network could be connected to each other (this would be a density of "1"). A density of "0.1" indicates that 10 % of these possible links exist within a network; the higher the number of nodes, the higher of course the number of possible links. Thus, in general, density tends to decrease with the size of a network. Therefore, it only makes sense to compare the densities of networks of (almost) the same size. Density can be interpreted as one indicator for the relative "cohesion", but also for the "complexity" of a network.[22] Other measurements are based on the equal or unequal distribution of quantitative characteristics such as degree among nodes; a high "degree centralisation" indicates that many links are concentrated on a relatively small number of nodes.[23] These distributions can also be statistically analysed and visualised for all nodes (by counting the frequency of single degree values) and used for the comparison of networks; again, certain distribution patterns (most prominently, power laws) are interpreted as "signatures of complexity" of a network ("scale free-networks").[24]

Networks are of course dynamic: relationships may be established, maintained, modified or terminated; nodes appear in a network and disappear (also from the sources). Standard tools of network analysis (still) force us to integrate these changes into one more or less static model. The common solution to capture at least part of these dynamics is to define "time-slices" (divided through meaningful caesurae in the development of the object of research, as defined by the researcher knowing the material) and to model distinct networks for each of them (see a simple example with two time slices in the following sub-chapter).[25]

---

[21] de Nooy – Mrvar – Batagelj, Exploratory Social Network Analysis 125-131. Prell, Social Network Analysis 171-172. For the small world-model cf. D. J. Watts, Small Worlds. The Dynamics of Networks between Order and Randomness (Princeton Studies in Complexity). Princeton – Oxford 1999; Sinha et al., Econophysics 212-216. See also Malkin, A Small Greek World.
[22] Prell, Social Network Analysis 166-168; Kadushin, Understanding Social Networks 29.
[23] Prell, Social Network Analysis 168-170.
[24] Newman, Networks 243-261; Sinha et al., Econophysics 208, 216-220; Fuhse, Soziale Netzwerke 103-105.
[25] de Nooy – Mrvar – Batagelj, Exploratory Social Network Analysis 92-95. Lemercier, Formale Methoden der Netzwerkanalyse 28-29; V. Batagelj – P. Doreian – A. Ferligoj – N. Kejžar, Understanding Large Temporal Networks and Spatial Networks. Exploration, Pattern Searching, Visualization and Network Evolution. Chichester 2014.

## 1.2 Example: a riverine transport network from Roman Antiquity to the Early Middles Ages

Within the project "Studies of inland harbours in the Frankish-German Empire as hubs for European communication networks (500-1250)", Lukas Werther (University of Jena) has created a database of harbours and landing sites at rivers and lakes in Central Europe, France and Northern Italy from the Roman period to the year 1000, also integrating the recently published catalogue of Christina Wawrzinek.[26] For a forthcoming paper, we use this data for the modelling and analysis of riverine traffic networks and a comparison of their structure in the Roman ($1^{st}$-$5^{th}$ cent. CE) and post-Roman ($6^{th}$-$9^{th}$ cent. CE) period.[27] In our model, harbour sites documented in historical or archaeological evidence serve as nodes and routes between them via river or lake navigation as links. Links in this model are both weighted (meaning that a quantity is attributed to them) and directed (a link leads from point A to point B, for instance). The aim is to integrate aspects of what Leif Isaksen has called "transport friction" into our calculations; otherwise, the actual costs of communication and exchange between sites, which would have influenced the frequency and strength of connections, would be ignored in network building. Links are weighted by using the inverted geographical distance between them; thus, a link would be the stronger the shorter the distance between two nodes ("distant decay" effect). Furthermore, directed links leading upstream (from point A to point B) are weighted with a third of the strength of links leading downstream (from point B to point A).[28]

---

[26] The project is part of the DFG-funded Priority-Programme 1630 "Harbours from the Roman period to the Middle Ages", see: http://www.spp-haefen.de/en/projects/binnenhaefen-im-fraenkisch-deutschen-reich/ resp. http://www.spp-haefen.de/en/home/. Cf. Ch. Wawrzinek, In portum navigare. Römische Häfen an Flüssen und Seen. Berlin 2014.

[27] J. Preiser-Kapeller – L. Werther, Connecting harbours. A comparison of traffic networks across ancient and medieval Europe (forthcoming paper), with beyond the Po catchment also includes models for the upper Danube, the Rhine and the Rhone.

[28] For the analysis of transport and traffic networks cf. J.-P. Rodrigue with Cl. Comtoi and B. Slack, The Geography of Transport Systems. 3rd ed., London – New York 2013, 307-317; E. J. Taaffe – H. L. Gauthier, Jr., Geography of Transportation. Englewood Cliffs, N. J. 1973, 100-158; C. Ducruet – F. Zaidi, Maritime constellations: A complex network approach to shipping and ports. *Maritime Policy and Management* 39, 2 (2012) 151-168. For a more general approach see M. Barthélemy, Spatial Networks. *Physics Reports* 499 (2011) 1-101. For transport networks of the past see: F. W. Carter, An Analysis of the Medieval Serbian Oecumene: A Theoretical Approach. *Geografiska Annaler*. Series B, Human Geography, Vol. 51, No. 1 (1969) 39-56; F. R. Pitts, The Medieval River Trade Network of Russia Revisited. *Social Networks* 1 (1978) 285-292; L. J. Gorenflo – Th. L. Bell, Network Analysis and the Study of past regional Organization, in: Ch. D. Trombold (ed.), Ancient road networks and settlement hierarchies in the New World. Cambridge 1991, 80-98; L. Isaksen, The Application of Network Analysis to Ancient Transport Geography: A Case Study of Roman Baetica, Digital Medievalist, 2008: http://www.digitalmedievalist.org/journal/4/Isaksen/; G. Graßhoff – F. Mittenhuber (eds.), Untersuchungen zum Stadiasmos von Patara. Modellierung und Analyse eines antiken geographischen Streckennetzes. Bern 2009. See also J. Leidwanger – C. Knappett et al., A manifesto for the study of ancient Mediterranean maritime networks (2014), online: http://journal.antiquity.ac.uk/projgall/leidwanger342. For transport friction cf. also R. J. van Lanen et al., Best travel options: Modelling Roman and early-medieval routes in the Netherlands using a multi-proxy approach. *Journal of Archaeological Science: Reports* 3 (2015) 144–159.

We modelled two networks, one on the basis of data for the 1st-5th cent. CE (period I) and one for the 6th-9th cent. CE (period II) and determined the standard centrality measures on the level of individual nodes (degree and especially betweenness and closeness) and of the entire network. While a visualisation of the nodes (sized according to their relative centrality) on a geographical map illustrates continuities and changes with regard to the focal points of connectivity in these two models **(see fig. 1-4)**, a comparison of quantitative measures indicates significant differences in range, connectivity and complexity between the two graphs **(see fig. 5)**: the number of nodes is 25 % smaller in the period II-network, the number of links is only half the one of the period I-model. Period II-network´s (weighted) density is only two thirds and its clustering coefficient half the size as for the period I-model. Significant is also the difference of values for transitivity (which indicates the percentage of link pairs in the network where when node A is linked to node B and node B is linked to node C also node A is linked to node C) between period I (0.657) and period II (0.25). Another measure especially developed for transport networks (as planar graphs) is circuitry, measuring the share of the maximum number of cycles or circuits (= a finite, closed path in which the initial node of the linkage sequence coincides with the terminal node) actually present in a traffic network model and thus indicating the existence of additional or alternative paths between nodes in the network and its relative connectivity and complexity[29]; here the difference is even more significant (0.38 for the period I-model and 0.10 for the period II-model). The model for the riverine network in period II is thus not only spatially more confined, but also less well connected and complex when compared with the model for period I **(see also fig. 6)**. This would correlate with assumptions on the relative reduction of organisational, economic and infrastructural complexity from the Roman to the post-Roman period.[30]

## 2 The complex network of the Roman Empire: a macro-perspective
### 2.1 "Complex" debates on the Roman Economy

It is safe to say that there does not exist consensus on core characteristics of "complexity" of the Roman imperial economy. One intense debate focuses on the degree of economic integration within the Roman Empire: was it an "enormous conglomeration of interdependent markets", whose degree of economic integration resulted also in the interdependence of prices in different regions (Peter Temin) or do we have to assume that "connectivity and isolation were unevenly spread across" a highly fragmented Mediterranean world with only some pockets of integrated markets (Paul Erdkamp, cf. also Peter Fibinger Bang)?[31] Another discussion centres on the role and share of the imperial state in the economy: was Rome a "tributary empire", whose transfers of goods for the army or the provision of the imperial capitals represented the predominant (or even only) sector of large-quantity trade? Did the empire´s demands and logistics at least very much determine orientation and scale of the axes of over-regional distribution and exchange both for the public and the private sector? Or is there an "overvaluation of the state-controlled economic sector" (Jean-Michel Carrié), leading to the "inadequate and somehow unrealistic idea that the imperial economy was controlled by a large redistributive system" (Bang).[32]

---

[29] Rodrigue et al., The Geography of Transport Systems 310, 313 and 315-316; Taaffe – Gauthier, Geography of Transportation 104-105: the circuitry or alpha-index is calculated as share of the maximum number of circuits actually present in a graph.
[30] For aspects of environmental change in this region cf. S. Cremonini – D. Labate – R. Curina, The late-antiquity environmental crisis in Emilia region (Po river plain, Northern Italy): Geoarchaeological evidence and paleoclimatic considerations. *Quaternary International* 316 (2013) 162-178.

As Jean-Michel Carrié has outlined, for sure there is an "overrepresentation of the (state) sector in surviving documents"[33]; but for our question, these sources at least provide evidence for considerations on the (minimum) scale and degree of infrastructural and organisational complexity necessary to maintain the "particular flow of resources and population directed by the imperial center" on which its success and survival depended (what Sam White for the Ottoman case has called the "imperial ecology").[34] When Emperor Julian in 362 CE provided 420,000 *modii* of wheat from imperial estates around the cities of Chalkis, Epiphania and Hierapolis for the starving population of the megalopolis of Antioch (ca. 160 km on the road west of Hierapolis), we may assume (according to Michael Decker) that "Julian mobilized the produce of more than 26,250 *iugera* of land and the sweat of more than 2,500 cultivators" in addition to 28,000 camels (with drivers, for the transport over land) for this supply "sufficient to feed approximately 262,500 adult males for a month or 4,468 families for a year". But we also learn that the first measure of the emperor had been the fixing of grain prices, which had provoked the major producers and dealers to "held their grain back from the market" for lack of acceptable profits. Only then the imperial apparatus had to step in.[35] In any case, this episode would qualify as sign of complexity of a system of various "economic agents (…) changing their actions and strategies in response to the outcome they mutually create", showing "a complicated mix of ordered and disordered behaviour" as proposed by Arthur and other theoreticians.[36]

---

[31] P. Temin, The Roman Market Economy. Princeton – Oxford 2013; P. Erdkamp, The Grain Market in the Roman Empire: A social, political and economic study. Cambridge 2005; P. F. Bang, The Roman Bazaar: A Comparative Study of Trade and Markets in a Tributary Empire. Cambridge 2008. Cf. also Bowman – Wilson, Quantifying the Roman Economy 15-28, and Jongman, Re-constructing the Roman economy, on various aspects of economic integration.
[32] Cf. the overview of positions in: J.-M. Carrié, Were Late Roman and Byzantine Economies Market Economies? A Comparative Look at Historiography, in: C. Morrisson, Trade and Markets in Byzantium. Washington, D. C. 2012, 13-26, esp. 20-21; Bang, The Roman Bazaar 68-69.
[33] Carrié, Were Late Roman and Byzantine Economies Market Economies 21.
[34] For the concept of imperial ecology and an analysis of the circuits of the imperial metabolism centred on Constantinople in the Ottoman period cf. S. White, The Climate of Rebellion in the Early Modern Ottoman Empire (*Studies in Environment and History*). Cambridge 2011, esp. 16-51 (17 for the citation). On models of flow cf. also J. K. Davies, Linear and nonlinear flow models for ancient economies, in: J. G. Manning – I. Morris, The Ancient Economy. Evidence and Models. Stanford 2005, 127-156.

Furthermore, recent debates in economic history somehow "vindicate" the significance of the state for economic development – in many cases, state activity was "not a sufficient, (…) but necessary condition" for growth. Even warfare can be interpreted as important "economic activity", and a military commander could be regarded as an economic agent like others.[37] The collapse of the (Western) Roman Empire also provides *argumenta ex negativo* for the relevance of the Roman imperial framework for economic complexity and its trajectory in its absence. While there may be no consensus on the degree of market integration in the Roman economy, on the basis of archaeological evidence, it seems clear that one of its most remarkable features was the "widespread diffusion" of goods (as especially evident from pottery), "not only geographically (sometimes being transported over many hundreds of miles), but also socially (so that it reached not just the rich, but also the poor)".[38] According to Bryan Ward-Perkins, the "end of complexity", on the contrast, was indicated by the reduction of the lateral as well as vertical range of connectivity, so that "even in the few places, like Rome, were pottery imports and production remained exceptionally buoyant, the middle and lower markets for good-quality goods (…) had wholly disappeared". The "dismembering of the Roman state, and the ending of centuries of security, were the crucial factors in destroying the sophisticated economy of ancient times."[39] Such an interpretation of the end of the system, in turn, implies a considerable degree of interdependence in the centuries before, since otherwise its "collapse" would not have affected even relatively peripheral regions such as Roman Britain to such a dramatic degree. Instead, after the disappearance of the overarching imperial framework, an agglomeration of "isolated", maybe "self-sufficient" clusters or "small worlds" of settlements and regions could have (re-)appeared, whose (maybe only slightly reduced) welfare would have depended mainly on their internal socio-economic dynamics as it did before – obviously, this what not the case, and the fragments of the former system were alone less than their sum (as could be expected for a complex system).[40]

---

[35] M. Decker, Tilling the Hateful Earth. Agricultural Production and Trade in the Late Antique East. Oxford 2009, 83-84 (with sources) and 257 (where Michael Decker makes a case "to rethink the nature of overland trade", at least in the Roman Near East: "it was neither dominated by luxury goods, nor was it infrequent").

[36] Arthur, Complexity and the Economy 1; Johnson, Simply Complexity 15-16.

[37] P. Vries, State, economy and the Great Divergence. Great Britain and China, 1680s-1850s. London et al. 2015; M. Mazzucato, The Entrepreneurial State. Debunking Public versus Private Sector Myths. London 2013; N. A. M. Rodger, War as an Economic Activity in the "Long" Eighteenth Century. *International Journal of Maritime History* 22 (December 2010) 1-18. For the Roman case see also the considerations in J. Löffl, Die römische Expansion. Berlin 2011, 313-475 (especially for the provinces in the Alps and at the Danube, also on the basis of archaeological and experimental data), and 476-486 (for the "autonomy" of Roman commanders as "economic agents" regarding logistics and administration, for instance).

[38] B. Ward-Perkins, The Fall of Rome and the End of Civilization. Oxford 2005, 88.

[39] Ward-Perkins, The Fall of Rome 106-107, 133. Cf. also M. McCormick, Origins of the European Economy. Communications and Commerce AD 300-900. Cambridge 2001, 778 and 782-783. On the distribution patterns of Roman pottery see also A. W. Mees, Die Verbreitung von Terra Sigilatta aus den Manufakturen von Arezzo, Pisa, Lyon und La Graufesenque. Mainz 2011.

## 2.2 Modelling the imperial traffic system: the "ORBIS Stanford Geospatial Network Model of the Roman World"

Can we capture aspects of the discussed integration and disintegration of the Roman economic system with the help of structural models? The most exhaustive network model of sea- and land routes of the Imperium Romanum is the "ORBIS Stanford Geospatial Network Model of the Roman World", developed by Walter Scheidel and Elijah Meeks in order to estimate transport cost and spatial integration within the Roman Empire. ORBIS is based on a network of roads, river and sea routes (in total, 1104 links) between 678 nodes (places), weighted according to the cost of transport.[41] **(fig. 7)**. Since it is aiming at the entirety of the empire´s traffic system, it is less detailed on the regional and local level than network models for smaller areas (such as the one presented above for the river Po). We have corrected this data (especially with regard to the localisations of some places) and modified the network model so that the link between two nodes (places) is the stronger the smaller the costs of overcoming the distance between them is (with regard to the travel time, according to the calculations of the ORBIS-team[42]), thus reflecting the ease or difficulty of transport and mobility between two localities (see also above 1.2).

In a first step, we analysed the spatial and statistical distribution of measures of centrality among the nodes of the network. The number and accumulated strength of links of a node (its weighted degree[43]) is of course high, where many localities are connected among each other via short distances via the most convenient transport medium – the sea (such as in the

---

[40] For a similar interpretation of the effects of the Late Bronze Age "Collapse" in the Eastern Mediterranean see now E. H. Cline, 1177 B. C. The Year Civilization collapsed. Princeton – Oxford 2014, esp. 164-170. In general on collapses in complexity cf. Arthur, Complexity and the Economy 144-157; J. A. Tainter, The Collapse of Complex Societies (*New Studies in Archaeology*). Cambridge 1988; J. A. Tainter, Social complexity and sustainability. *Ecological Complexity* 3 (2006) 91–103; M. Scheffer, Critical Transitions in Nature and Society (*Princeton Studies in Complexity*). Princeton 2009; R. T. T. Forman, Urban Ecology. Science of Cities. Cambridge 2014, esp. 65–90.
[41] W. Scheidel – E. Meek et al., ORBIS: The Stanford Geospatial Network Model of the Roman World: http://orbis.stanford.edu/. The data set was downloaded from: https://purl.stanford.edu/mn425tz9757 (Creative Commons Attribution 3.0 Unported License); Authors of the data set are E. Meeks – W.Scheidel – J.Weiland – S. Arcenas. For a similar model see also Sh. Graham, Networks, agent-based models and the Antonine Itineraries: implications for Roman archaeology. *Journal of Mediterranean Archaeology* 19 (1) (2006) 45–64.
[42] In case of parallel links between nodes in the data set, the "cheapest" connection was selected.
[43] Newman, Networks 168-169.

Aegean). Statistically, the distribution of these degree values is very unequal, with a high number of nodes with relatively low degree centrality and a small number of "hubs" with high centrality values **(fig. 8)**. As indicated above, betweenness on the contrast measures the relative centrality of a node in the entire network due to its position on many (or few) potential shortest paths between nodes.[44] In the ORBIS-network, the hubs of maritime transport serve as most important integrators of the entire system in this regard; at the same, the statistical distribution of betweenness values is even more unequal than the one of degree centrality **(fig. 9)**. "Closeness" in turn measures the average length of all paths between a node and all other nodes in a network and indicates its overall centrality (or remoteness).[45] Statistically, closeness-values are relatively equally distributed; but their spatial distribution demonstrates the decisive role of maritime connectivity via the Mediterranean for the cohesion of the entire network **(fig. 10)**. The ORBIS-model, which is also characterised by a (relatively to its size) high value of circuitry (0.32, see above 1.2 for this measure), thus shows several "signatures of complexity" of large scale networks.

The ORBIS-model can also be used to approach structural differentiations within the Roman traffic system; as outline above, networks are often structured in clusters, meaning groups of nodes which are more densely and closely connected among each other than with the rest of the network. For the identification, we use the algorithm for "group detection" developed by M. Newman, which aims at an "optimal" partition of the network into clusters.[46] Complex network are characterised by "nested clustering", such that within clusters further sub-clusters can be detected, within which further cluster can be identified, across several levels of hierarchy.[47]

With the help of the Newman-algorithm we identify 25 regional resp. over-regional clusters of higher internal connectivity within the ORBIS-model **(fig. 11)**:

| Newman-Cluster nr. | Regions |
| --- | --- |
| 1 | Upper Danube, Eastern Alps |
| 2 | North Syria, Northwest Mesopotamia, South Asia Minor |
| 3 | Rhine area |
| 4 | Middle Danube, North Balkans |
| 5 | North and central Adriatic |

---

[44] Newman, Networks 185-193.
[45] Wassermann – Faust, Social Network Analysis 184-188.
[46] Newman, Networks 372-382.
[47] Barabási, Network Science 331-338.

| 6 | Central North Africa, East coast of Iberian Peninsula, Baleares |
|---|---|
| 7 | South Iberian Peninsula, West North Africa |
| 8 | Region around the Sea of Marmara, North Aegean |
| 9 | Central and Northwest Aegean |
| 10 | Central South Italy |
| 11 | Palestine |
| 12 | Britannia and Channel coast |
| 13 | Rome, Latium and Campania |
| 14 | Egypt |
| 15 | Cyrenaica, Crete and South Peloponnese |
| 16 | Southwest Asia Minor |
| 17 | Cyprus and north coasts of Levant |
| 18 | East North Africa, Sicily and Southwest of South Italy |
| 19 | Black Sea and North of Asia Minor |
| 20 | Etruria, Liguria, Corsica and Southeast of Gaul |
| 21 | Gaul and North of the Iberian Peninsula |
| 22 | South Adriatic and North Epirus |
| 23 | Western plain of the river Po |
| 24 | Northwest central Greece, North Peloponnese and Ionian Sea |
| 25 | Central Aegean (micro-cluster) |

The majority of these clusters owe their connectivity again to either maritime connections (nr. 5, 6, 7, 8, 9, 12, 13, 15, 16, 17, 18, 19, 20, 21, 22, 24, 25) or riverine routes (nr. 1, 3, 4, 14, 23).[48] In order to test the concept of "nested clustering", we applied the Newman-algorithm also on each of the 25 (over)regional clusters, resulting in the identification of between three and eight local or regional sub-clusters within each of the larger clusters (**fig. 11**). This complex network model of localities and routes in the Roman Empire therefore across several spatial scales can be perceived as a system of nested clusters, down to the level of individual settlements and their hinterlands. In such a network, speed and cohesion of empire-wide connectivity depends on the trans-regional links between these clusters which structure the entire system. It is therefore also somehow located between the scenario of a fragmented

---

[48] See also McCormick, Origins of the European Economy 77-114, on the significance of riverine and maritime shipping.

Roman Mediterranean of Erdkamp of Bang and the model of a Roman Empire economically integrated by trade links (Temin, see above).

But what happens, if these relatively cost-intensive, "fragile links between different people and different economies" across larger distances, as Ward-Perkins calls them[49], "disappear"? In order to target this question, we eliminated step by step all links from the model which would "cost" more than five, more than three, more than two and finally more than one day´s journey(s) (according to the calculations of the ORBIS-team) (**fig. 12**). The result is an increasing fragmentation of the network in components of different size, partially along the "fault lines" between the clusters and sub-clusters, which we identified for the unmodified network model. But even if we eliminate the connections across longer distances, some larger, over-regional clusters especially of maritime connectivity demonstrate remarkable "robustness".[50] In the model, in which all connections which "cost" more than one day´s journey are deleted, the largest still fully connected component is located in the Eastern Mediterranean between the Tyrrhenian Sea and the Levant, with its centre in the Aegean (**fig. 12**). This would correspond to the central regions and communication routes which remained under control of the (Eastern) Roman Empire after the loss its eastern provinces to the Arabs in the 7th century, at the end of an actual process of increasing fragmentation of the (post)Roman world.[51] Yet besides the resilience of maritime connectivity in regions of the Eastern Mediterranean (and an uninterrupted cohesion of the "Egyptian" cluster[52]), we observe a general "disentanglement" of large parts of the Roman traffic system, especially in the West of Europe, equally in the interior of the Balkans or also between the North and Souths coasts of the Mediterranean. The model is of course no perfect depiction of historical reality, but at best an appropriation towards certain structural parameters of the web of transport links within the Imperium Romanum. Nevertheless, we observe some remarkable parallels to actual historical processes of the 5th to 7th century CE (Chris Wickham for instance wrote about a partial „micro-regionalisation" of the „Mediterranean world-system" during this period)[53], which hint at the impact of processes of integration respectively disentanglement especially due to the establishment and growth respectively the contraction of long distance connections.[54]

---

[49] Ward-Perkins, The Fall of Rome 382.
[50] Cf. McCormick, Origins of the European Economy 565-569 (with map 19.2) on the "resilience" of certain sea routes in the 7th to 9th cent. CE.
[51] L. Brubaker – J. Haldon, Byzantium in the Iconoclast Era c. 680-850: a History. Cambridge 2011.
[52] A further remarkable parallel to the actual economic development, cf. Ch. Wickham, Framing the Early Middle Ages. Europe and the Mediterranean, 400-800. Oxford 2005, 759-769.
[53] ickham, Framing the Early Middle Ages 693-794, esp. 778-780, 792-794; Ch. Wickham, The Mediterranean around 800: On the Brink of the Second Trade Cycle. *Dumbarton Oaks Papers* 58 (2004) 161-174; McCormick, Origins of the European Economy 28-63. See also Ward-Perkins, The Fall of Rome.

**3 A network of places and commodities on the basis of one piece of textual evidence**

The traffic system of the Roman Empire (for which a model was just presented) served as infrastructure for the mobility of humans, the transport of commodities and thus any form of market exchange based both on commercial and non-commercial activities (on the discussion of respective shares of these segments in the Roman economy, see above). One of the most interesting contemporaneous texts in this regard does not touch upon the Mediterranean centre of the Imperium, but on its foreign trade: the "Periplus of the Erythraean Sea", a guide to trade and navigation in the Indian Ocean usually dated to the 1st century CE.[55] Most recently, Eivind Heldaas Seland has used this text as basis for the modelling of various networks, demonstrating also the successful application of the method on a single (albeit also unique) piece of evidence for questions of Roman economic history; as he explains: "first, the text describes existing networks of people, places and commodities at the time of its composition. Second, the text allows us speculate on possible and potential linkages that are not definitively described. Third, the text itself can be approached as an inclusive macro-network where words, for instance those describing places, relate to other words de-scribing products. It is this latter aspect of the textually conceived network that allows us to reconstruct former networks that were actually in existence or might well have been so."[56] Especially the last approach, via which Seland models a two-mode network of localities and of goods either exported from or imported to these places, provides a most interesting insight into the complexity of circuits of ancient exchange in the Indian Ocean. Based on the data set provided generously by Seland online[57], we were able to rebuild this network of 39 places and 112 commodities to apply further manipulations and analyses on it (**fig. 13**). As Seland demonstrates, "the advantage of this network is that it allows us to look at supply/demand relationship in first-century Indian Ocean trade. While the narrative of the Periplus relates only what the author knew was traded in each port, the graph gives access to information on all the places where these products were available."[58] For further analysis, on the basis of this two-mode (or "affiliation") network I modelled two one-mode-networks: one of commodities, where two commodities are connected if they have at least one marketplace in common **(see fig. 14)**, and one of marketplaces, where two localities are connected if they have at least one commodity in common **(see fig. 16)**. Both networks are weighted networks, with the strength of links differing according to the number of common marketplaces or commodities. The network of commodities reflects the co-occurrence of items on the same marketplaces and illustrates their relative ubiquity or special positions within the web of exchanges (as narrated in the text). The network of marketplaces reflects the relative similarity of markets with regard to the range and variety of commodities traded there (and not links of direct exchange, as is a frequent misunderstanding especially if such "affiliation-networks", also often used to analyse the co-occurrence of artefact types on archaeological sites, are visualised on a geographical map**, see fig. 17**).[59]

---

[54] The transport of larger amounts of commodities and numbers of people as common in the Roman imperial framework could not be compensated to a comparable amount which would have guaranteed the enduring cohesion of the Mediterranean system by new forms of mobility such as pilgrimage to the Holy Land or the transfer of relics which were continued during and after the crisis of Late Antiquity, cf. also McCormick, Origins of the European Economy 270-277, 385-387.
[55] L. Casson, The Periplus Maris Erythraei: Text with Introduction, Translation, and Commentary. Princeton 1989. Cf. also G. Parker, The Making of Roman India (Greek Culture in the Roman World). Cambridge 2008.
[56] E. H. Seland, The Periplus of the Erythraean Sea: A Network Approach. *Asian Review of World Histories* 4:2 (July 2016) 191-205, esp. 194.
[57] Permalink to dataset at: http://bora.uib.no/handle/1956/11470 (Dataset published under Creative Commons license 4.0: http://creativecommons.org/licenses/by/4.0).

The network of commodities consists of 112 nodes and 2135 links and is the most complex model (in terms of the number of connections) presented in this paper. The values for density (0.34), the clustering coefficient (0.79) and transitivity (0.71) suggest a relatively high ubiquity of commodities among places so that the maximum distance between two nodes is 2.24 (a "small world" of goods[60]). Still, an analysis of the actual distribution of weighted degree values among commodities shows a high inequality in the accumulated strength of ties of individual nodes **(see fig. 14)**.[61] The heavy weights of relative ubiquity with the highest degree values are grain, wine, [Roman] money, tin and slaves. The first four of these goods are also the only ones which connect all the places with the highest degree values in the core of similarity among places. The highest betweenness values, on the contrast, are those of silverware, *molochinon* (from which cloth and garments were produced[62]), cotton-garments, frankincense and precious stones; these more luxurious products co-occur with goods otherwise not to be found in the same circuits of distribution and serve as "intermediaries" between these circuits in the network (**fig. 15**).

---

[58] Seland, The Periplus of the Erythraean Sea 202.
[59] For this method and its application especially in archaeological network analysis cf. S. M. Sindbæk, Broken links and black boxes: material affiliations and contextual network synthesis in the Viking world, in: C. Knappett (ed.), Network-Analysis in Archaeology. New Approaches to Regional Interaction. Oxford 2013, 71-94; T. Brughmans, Thinking through networks: a review of formal network methods in archaeology. *Journal of Archaeological Method and Theory* 20 (2012) 623-662; P. Östborn – H. Gerding, Network analysis of archaeological data: a systematic approach. *Journal of Archaeological Science* 46 (2014) 75-88; J. Preiser-Kapeller, Thematic introduction, in: J. Preiser-Kapeller - F. Daim (eds.), Harbours and Maritime Networks as Complex Adaptive Systems (RGZM Tagungen). Mainz 2015, 1-24. For similarities between sites as basis for network modelling cf. also P. Östborn – H. Gerding, The Diffusion of Fired Bricks in Hellenistic Europe: A Similarity Network Analysis, in: A. Collar – F. Coward – T. Brughmans – B. J. Mills, The Connected Past: critical and innovative approaches to networks in archaeology. A special issue of the *Journal of Archaeological Method and Theory* 22 (1) (2015) 306-344.
[60] See above on the notion of the "small world network".
[61] On such distribution patterns in economics cf. Sinha et al., Econophysics 115-123.

Also the one-mode-network of the 39 places connected through ties of co-occurrence of commodities seems to be a densely interwoven "small world" with an average path length of two, a density of 0.35 and a clustering coefficient of 0.732 (**fig. 16**). But nodes are again unequally integrated into this network; the accumulated weighted degree values of the top nine nodes (Myos Hormos, Berenike, Barygaza, Muziris and ex aequo Nelkynda, Kamara, Poduke and Sopatma) amount to 51 % of the total, with the Egyptian harbours of Myos Hormos and Berenike alone representing 16 %. Between these two places can be found also the strongest link of similarity (tie strength 64), while the next strongest tie (between Mundu and Mosyllon) amounts only to 11. The top five nodes in betweenness centrality, however, are located on either side of the pivotal Bab-el-Mandeb between East Africa and the Arab peninsula (Avalites, Muza) or in India (Barygaza, Ozene, Taprobane) (**fig. 17**); only then, Myos Hormos shows up in the list. From the view of the entire network, the Egyptian harbours are important "players", but also located at the Western periphery of the overall exchange system, whose integration depends on other intermediary nodes.

Our analysis thus confirms the findings of Seland: the bias of the Periplus towards the perspective of traders coming from Roman Egypt and aiming at exchanging their products for those provided elsewhere makes itself clearly felt also in the network model. Yet also "Arabian, Indian, Persian Gulf and Bay of Bengal circuits" and the centrality of other nodes "become more visible" by such an "exercise"; network analysis helps to extract this implicit information, which is embedded in the text, but can be identified through reading only with difficulties.[63] We can therefore approach in a different way the structural and commercial parameters under which Roman trade into the Indian Ocean was entangled with various regional and over-regional circuits (this clustering also becomes visible if we apply the Newman-algorithm on the network of places, see **fig 18**), summing up to another complex commercial system beyond the *Mare Internum* of the Mediterranean.[64]

---

[62] Casson, The Periplus Maris Erythraei 249, assumes that these were also cotton garments of especially high quality, but debate is still open; cf. also Parker, The Making of Roman India 157.
[63] Seland, The Periplus of the Erythraean Sea, esp. 204-205.
[64] This structuring of a complex web of commodities and markets, inherent in the information stemming from only one source, shows similarities with results from the application of network theory to modern-day data on the combination of countries and products that they export by Ricardo Hausmann and César A. Hidalgo; on the basis of a two-mode-network-model respectively its transformation into a "network of relatedness between products", defining a "product space of world economy", they were able to detect subtle differences in the

## 4 Micro-perspectives and qualitative approaches of network analysis

After inspecting network models of river ports, routes, commodities and marketplaces we may remember Brian´s statement from the beginning of this paper that complexity economy is based on the assumption that "economic agents (…) constantly change their actions and strategies in response to the outcome they mutually create" and ask: where are these agents? Of course, we assume that network structures, (changing) relative positions of nodes or distribution patterns emerge from the interplay of these agents (be it the emperor, a merchant, a craftsman, a peasant or the associations and organisations they form[65]). The paper of Xavier Rubio Campillo and colleagues prepared for this volume, for instance, comes to the conclusion (based on the stamps from amphorae in Monte Testaccio) that "olive oil production was structured similarly as current firm-size distributions (i.e. it follows a power law)" and supports our assumption on emergent complex properties.

To reflect on the actual social interactions behind this statistical pattern, it may be helpful to take into consideration Harrison White´s elaborate model for markets as networks of firms; he perceives "markets are tangible cliques of producers watching each other" (and less the consumers), creating an emerging "pecking order" of firms. This hierarchy becomes "taken for granted" in the form of a "self-reproducing role structure of relations among the producers". Critics have observed that such dynamics are imaginable only for small markets with maybe a handful of producers; thus, this model could be more valid for pre-modern than modern conditions.[66] But while late medieval material allows us to survey, visualise and model the actual networks of interactions between economic agents[67], we lack in most cases

---

positions of countries regarding the ubiquity or diversity of their products, also correlating with their relative economic performance through time, cf. C. A. Hidalgo – R. Hausmann, The building blocks of economic complexity. *Proceedings of the National Academy of Sciences of the United States of America* 106 no. 26, 10570–10575; R. Hausmann – C. A. Hidalgo, The Network Structure of Economic Output. *Journal of Economic Growth* 16/4 (December 2011) 309-342; Sinha et al., Econophysics 230-234. See also now the "popularised" version of these findings in: C. Hidalgo, Why information grows. The Evolution of Order, from Atoms to Economies. London 2015, 129-142, and the online-Atlas of Economic Complexity: http://atlas.cid.harvard.edu/. For a further development of this approach see G. Caldarelli et al., A Network Analysis of Countries´ Export Flows: firm Grounds for the Building Blocks of the Economy. *PLOS One* 7/10 (October 2012) (http://journals.plos.org/plosone/article?id=10.1371/journal.pone.0047278), and M. Cristelli – A. Tacchella – L. Pietronero, The Heterogeneous Dynamics of Economic Complexity. *PLOS One* 10/2 (February 2015) (http://journals.plos.org/plosone/article?id=10.1371/journal.pone.0117174).

[65] Cf. for instance T. T. Terpstra, Trading Communities in the Roman World. A Micro-Economic and Institutional Perspective. Leiden – Boston 2013.

[66] H. White, Markets from Networks: Socioeconomic models of production. Princeton 2002; cf. also Knoke, Economic Networks 60-64.

[67] J. F. Padgett – P. D. McLean, Organizational Invention and Elite Transformation: The Birth of Partnership Systems in Renaissance Florence. *American Journal of Sociology* 111/5 (March 2006) 1463–1568; M. Burkhardt, Der hansische Bergenhandel im Spätmittelalter: Handel – Kaufleute – Netzwerke. Vienna – Cologne

evidence of comparable density for earlier periods of Mediterranean history but again have to rely on "proxies". Similar to Campillo´s work, Shawn Grahams analysis of the "network dynamics of the Tiber Valley Brick Industry" relies on the co-occurrence of stamps on various sites and highlights some structural dynamics of these networks (with changing degree distribution patterns reflecting different organisational patterns, for instance).[68] Based on the data collected for the Roman potter shops (of *terra sigillata*) of Rheinzabern (Tabernae) by Allard W. Mees, I prepared a similar network model for the connections between potters respectively potter groups due to the co-occurrence of commonly used hallmarks on their widely distributed products. Also here, one observes an highly unequal distribution of (weighted) degree values for the network model across the entire period of activity of Rheinzabern (ca. 150-270 CE) (**fig. 19**), but at the same time differences in the density and structure of connectivity between the eight potter groups Mees has identified, whose activities started and ended at different points in time and targeted different sales areas **(see fig. 20 and 21)**.[69] These differences therefore could also reflect different forms of inner organisation or strategies of cooperation, with some groups depending on close interaction and stronger centralisation, while others were more loosely structured.

The emergence of new production sites of *terra sigillata* in various provinces is also one example for the spread of skills and technology via networks between places and agents; such phenomena of diffusion have been one focus of dynamic network modelling, especially regarding the dependence of diffusion patterns on underlying network structures. By inference, such patterns in turn provide clues on the density, structure and axes of interaction of networks, as Lars Boerner and Battista Severgnini have demonstrated by using the spread of the Black Death as a proxy to measure economic interactions in 14[th] century Europe, for instance. For the 6[th]-8[th] century, diffusion patterns of the waves of the "Justinianic Plague" could provide similar insights, but unfortunately the density of evidence is much smaller.[70] In

---

2009; F. Apellániz, Venetian Trading Networks in the Medieval Mediterranean. *Journal of Interdisciplinary History* 44/2 (2013) 157-179; J. Preiser-Kapeller, Liquid Frontiers. A relational analysis of maritime Asia Minor as religious contact zone in the 13th-15th century, in: A. Peacock et al. (eds.), Islam and Christianity in Medieval Anatolia. Aldershot 2015, 117-146.

[68] Sh. Graham, Ex Figlinis: The Network Dynamics of the Tiber Valley Brick Industry in the Hinterland of Rome (*BAR International Series*). Oxford 2006. Cf. also now Sh. Graham – S. Weingart, The Equifinality of Archaeological Networks: an Agent-Based Exploratory Lab Approach, in: A. Collar – F. Coward – T. Brughmans – B. J. Mills, The Connected Past: critical and innovative approaches to networks in archaeology. A special issue of the *Journal of Archaeological Method and Theory* 22 (1) (2015) 248-274.

[69] Cf. A. W. Mees, Organisationsformen römischer Töpfer-Manufakturen am Beispiel von Arezzo und Rheinzabern. 2 vol.s, Mainz 2002.

[70] Cf. Mees, Die Verbreitung von Terra Sigilatta; Jackson, Social and Economic Networks 185-122, and Easly – Kleinberg, Networks, Crowds, and Markets 497-604 (for modes of diffusion on networks). L. Boerner – B. Severgnini, Epidemic Trade. London School of Economics, *Economic History Working Papers* No: 212/2014.

any case, both skills and germs depended on the mobility of individuals to spread, which leads back to the question of networks of economic agents and groups. In his recent book on "trading communities" in the Roman Empire, Taco T. Terpstra stated that "information circulated within small but far-reaching groups defined by their members' shared geographical origin; the loss of a member's reputation or trading position within the group formed the instrument of contract enforcement." As Terpstra himself makes clear, he borrowed heavily from the findings on medieval trading communities (or "diasporas") such as the studies of Avner Greif.[71] Eivind Heldaas Seland was able to propose some network models for the diffusion of and connections between trading communities (such as those from Palmyra) in the ancient Western Indian Ocean.[72] But again, only the dense evidence for late medieval trade diasporas would allow for a more elaborate structural and quantitative analysis.[73] These communities highlight likewise the significance of the "overlap" of various qualities and categories of ties – common geographic origin, ethnic or religious affiliation, kinship and economic exchange – for the emergence of more durable forms of partnership and association; John F. Padgett has analysed the "birth of Partnership Systems in Renaissance Florence" with the help of a model of "multiple-network ensembles".[74]

The diffusion of technological skills or agricultural practices implies also connections not only between people, but also with and between plants, animals and objects. One could reflect on the socio-economic agency of the silk worm, reportedly smuggled by Christian monks from China to Byzantium in the time of Emperor Justinian I (527-565), or of the sugar cane, allegedly brought from India to China by Buddhist monks in the 7th century CE; they imported also the complex and intensive entanglements connected with the breeding and

---

For an analysis of the diffusion of religious ideas in the Roman Empire with the help of network theory cf. A. Collar, Religious Networks in the Roman Empire. The Spread of New Ideas. Cambridge 2013.

[71] Terpstra, Trading Communities in the Roman World 2. A. Greif, Institutions and the Path to the Modern Economy: Lessons from Medieval Trade. Cambridge 2006. But see also Sh. Ogilvie, Institutions and European Trade. Merchant Guilds, 1000-1800. Cambridge 2011, with some critical evaluation of Greif´s theses.

[72] E. H. Seland, Networks and social cohesion in ancient Indian Ocean trade: geography, ethnicity, religion. *Journal of Global History* 8, Issue 03 (November 2013) 373-390; E. H. Seland, Trade and Christianity in the Indian Ocean during Late Antiquity. *Journal of Late Antiquity* 5, 1 (Spring 2012) 72-86. For some considerations on the interplay between the diffusion of trading or religious diasporas and economic connectivity in that period cf. also J. Preiser-Kapeller, Peaches to Samarkand. Long distance-connectivity, small worlds and socio-cultural dynamics across Afro-Eurasia, 300-800 CE. Working paper for the workshop: "Linking the Mediterranean. Regional and Trans-Regional Interactions in Times of Fragmentation (300 -800 CE)", Vienna, 11th-13th December 2014 (online: http://oeaw.academia.edu/JohannesPreiserKapeller/Papers).

[73] Cf. Apellániz, Venetian Trading Networks; J. Preiser-Kapeller, Un)friendly takeover. A comparison between the emergence of "Western" maritime networks in the "East" in the Mediterranean and the Indian Ocean, 800-1700 CE (forthcoming paper).

[74] J. F. Padgett – W. W. Powell, The Emergence of Organizations and Markets. Princeton – Oxford 2012, 168-207, and 70-114 (for the wider theoretical framework on "autocatalytic networks", inspired by findings from chemistry, for the emergence of organisations and markets).

manufacture of these products, creating "communities of practice" as well as new networks between producers, traders and consumers across large distances.[75] Theorists of Actor-Network-Theory (ANT) such as Bruno Latour postulate to regard humans and objects as equal actors within a network; he states: "anything that modifies a state of affairs by making a difference is an actor (…). Thus, the question to ask about any agent is simply the following: does it make a difference in the course of some other agent´s action or not?[76] ANT has found some attention in archaeology, especially in two books by Carl Knappett and Ian Hodder.[77] The latter emphasises the intensity of entanglements between humans and things: "things depend on people when they are procured, manufactured, exchanged, used and discarded but in particular they depend on people to maintain them if they are to remain as people want them. Or they depend on humans to maintain the environments in which they thrive. Made things are not inert or isolated. Their connections with other things and their maintenance depend on humans. (…) this dependence of things on humans draws humans deeper into the orbit of things. Looking after things as they get depleted or fall apart or as they grow and reproduce trap humans into harder labor, greater social debts and duties, changes schedules and temporalities. (…) Humans have had increasingly to invest labor and new technologies to manage and sustain these things and have found themselves organized by them."[78]

Again, this hints at the actual degree of (inevitable) complexity and the number and scale of feedbacks inherent in most networks of economic agency we may encounter in the Roman or pre-modern history in general (already among the Neolithic farmers of Çatalhöyük, which Hodder uses as examples in his book).[79] Latour pleads for a complete survey of all possible entanglements within a site and beyond across spatial and temporal scales with all actors and places necessary for a specific place to do something ("localising the global, globalising the local").[80] At the same time, we have to be aware of aspects of bias, selection, manipulation and fragmentariness inherent in all our pieces of evidence, be it an archaeological assemblage or a text. But we could understand all these phenomena as "narratives of entanglements", which provide us with a certain perspective, a specific extract of the actual totality of entanglements (impossible to capture even for modern-day cases).[81] Both Network Theory and Narratology lead us again to the possibility of quantitative and structural analysis – and in the case of "Quantitative Narrative Analysis" as developed by Roberto Franzosi, they are flowing together. Franzosi wrote: "Narrative texts are doubly relational. They depict both social relations and conceptual relations. (…) It is one thing to be able to say which (and perhaps how often) themes, concepts, actors appear in a text and another to be able to map the network of relations that give meaning to a text (or the social world)." Thus, he applied tools of quantitative network analysis to map the relations of violence between actors in the narratives of Italian newspaper in the period of the rise of Fascism.[82] The example of Seland´s analysis of the Periplus of the Erythrean Sea (see above), but also other recent studies illustrate the potential of this approach for the analysis of historical texts.[83]

---

[75] For silk see A. Muthesius, Byzantine Silk Weaving AD 400 to AD 1200. Vienna 1997, esp. 5-26, for sugar: S. Mazumdar, Sugar and Society in China: Peasants, Technology, and the World Market. Harvard 1998; M. Ouerfelli, Le Sucre. Production, commercialisation et usages dans la Méditerranée médiévale. Leiden – Boston 2008. In general see also V. Roux, Spreading of Innovative Technical Traits and Cumulative Technical Evolution: Continuity or Discontinuity? *Journal of Archaeological Method and Theory* 20 (2013) 312–330 (with further literature).
[76] B. Latour, Nonhumans, in: St. Harrison – St. Pile – N. Thrift (eds.), Patterned Ground. Entanglements of Nature and Culture (London 2004) 224-227; cf. also B. Latour, Reassembling the Social. An Introduction to Actor-Network-Theory. Oxford 2005.
[77] C. Knappett, An Archaeology of Interaction. Network Perspectives on Material Culture and Society. Oxford 2011; I. Hodder, Entangled. An Archaeology of the Relationships between Humans and Things. Malden, Oxford 2012.
[78] Hodder, Entangled 85-87.
[79] For the most interesting Chinese case cf. also L. Leddertose, Ten Thousand Things. Module and Mass Production in Chinese Art. Princeton 2000.

Targeting the inherent complexity in narratives may provide even the opportunity to use "big data" for an analysis of the relations between places, individuals and objects in the ancient Mediterranean. Especially the project "Pelagios: Enable Linked Ancient Geodata In Open Systems" (by Leif Isaksen, Elton Barker and Rainer Simon) has demonstrated the potential of linking data from large collections of texts and information on sites and artefacts[84]; the platform provides also tools to map the spatial distribution of this data and the connections between places on the basis of narrative co-reference (Pelagios Graph Explorer) and demonstrates their application on an increasing number of classical (but recently also medieval) texts.[85] At the same time, modern language statistics techniques (such as Latent Semantic Analysis) focus on the analysis of such large text corpora; recent studies suggest that co-occurrences of named entities (city names and person names) in a big number of texts can be used to estimate the longitude and latitude of cities and the relations between places or between individuals.[86] Further developments of these methods may allow scholars to fully exploit the potential of the mass of evidence which is already there to analyse the complex relational webs framing respectively emerging from the interplay between (economic) agents in the Roman world.

---

[80] Latour, Reassembling the Social 173 and 200-202. Cf. also J. Preiser-Kapeller, The Maritime Mobility of Individuals and Objects: Networks and Entanglements, in: J. Preiser-Kapeller - F. Daim (eds.), Harbours and Maritime Networks as Complex Adaptive Systems (RGZM Tagungen). Mainz 2015, 119-140.

[81] Cf. also J. McGlade, Simulation as Narrative: Contingency, Dialogics, and the Modeling Conundrum. *Journal of Archaeological Method and Theory* 21 (2014) 288–305. For the possibility to use calculations on the input of energy and manpower into major building projects as proxies for the scale and complexity of economies and the framework of "energetics" for the Byzantine case see J. Pickett, Beyond Churches: Energetics and Economies of Construction in the Byzantine World (working paper, online: http://www.academia.edu/690185/_Energetics_and_Economies_of_Construction_in_the_Byzantine_World_).

[82] R. Franzosi, Quantitative Narrative Analysis. Los Angeles – London 2010; R. Franzosi, From Words to Numbers. Narrative, Data and Social Science. Cambridge 2004.

[83] Cf. also A. Crespo Solana (ed.), Spatio-Temporal Narratives: Historical GIS and the Study of Global Trading Networks (1500-1800). Cambridge 2014; R. Senturk, Narrative Social Structure: Anatomy of the Hadith Transmission Network, 610-1505. Stanford 2005; H. Fernández-Aceves, A Relational View of the Norman Kingdom of Sicily and its Royal Court: The Social Space Constructed by 'Hugo Falcandus'. Master-Thesis, Central European University Budapest 2013; J. Preiser-Kapeller, Entangling Maragha. Mapping and Quantifying the Networks of a Medieval Urban Centre, in: J. Pfeiffer (ed.), Maragha and its Scholars. The Intellectual Culture of Medieval Maragha, ca. 1250-1550 (forthcoming, pre-print online: https://oeaw.academia.edu/JohannesPreiserKapeller/Papers).

[84] http://pelagios-project.blogspot.co.at/. See also the related project "Google Ancient Places": https://googleancientplaces.wordpress.com/.

**5 Conclusion**

Network theory aims at central aspects of (economic) complexity: the entanglements between agents and the structural patterns of relations and connections framing and emerging from their interactions. It allows for a modelling of networks across scales (socially – from the individual to the level of cities and polities, spatially – from one production site up to an entire empire, and temporally – from static snapshots for periods of different duration to series of time slices) and for an overlap of webs of ties of different qualities and categories, also stemming from different sources ("multiplex networks"[87]). Elaborate structural and quantitative analysis depends on a considerable density of evidence, but as demonstrated a single text may suffice to provide a glimpse at the inherent complexity of past economic interaction; also thereby, network models can serve as proxies in order to estimate the range, density and complexity of past economic life. Equally, the relational approach invites to a structural and quantitative comparison between periods, regions and the economic systems of polities and empires.[88] An increasing number of proxies of this kind will allow us to capture the trajectories of economic complexity (beyond metaphors) from antiquity into the middle ages, may they be characterised by evolutionary dynamics, self-reinforcing processes or phase transitions towards system states of augmented or significantly reduced complexity. Their creation as well as interpretation demands an even more intensive dialogue between humanities and sciences.[89]

---

[85] See for instance E. Barker – St. Bouzarovski – Ch. Pelling – L. Isaksen, Writing space, living space. Time, agency and place relations in Herodotus's Histories, in: J. Heirman – J. Klooster (eds.), The Ideologies of Lived Space in Literary Texts, Ancient and Modern. Ghent 2013, 229–247.

[86] Cf. M. M. Louwerse – R. A. Zwaan, Language encodes geographical information. *Cognitive Science* 33 (2009) 51-73; M. M. Louwerse – St. Hutchinson – Z. Cai, The Chinese Route Argument: Predicting the Longitude and Latitude of Cities in China and the Middle East Using Statistical Linguistic Frequencies, in: Proceedings of the 34th Annual Conference of the Cognitive Science Society. Austin, Texas 2012, 695-700; St. Hutchinson – V. Datla – M. M. Louwerse, Social networks are encoded in language, in: Proceedings of the 34th Annual Conference of the Cognitive Science Society. Austin, Texas 2012, 491-496.

[87] For an impressive example see G. Earl – S. Keay – T. Brughmans, Complex Networks in Archaeology: Urban Connectivity in Iron Age and Roman Southern Spain, 2014 (Abstract online: http://eprints.soton.ac.uk/336993/1/brughmans_keay_earl_AHCN_Leonardo_v3.pdf; full publication in preparation). See also now T. Brughmans – S. Keay – G. Earl, Understanding Inter-settlement Visibility in Iron Age and Roman Southern Spain with Exponential Random Graph Models for Visibility Networks, in: A. Collar – F. Coward – T. Brughmans – B. J. Mills, The Connected Past: critical and innovative approaches to networks in archaeology. A special issue of the *Journal of Archaeological Method and Theory* 22 (1) (2015) 58-153. Cf. also J. Preiser-Kapeller, Networks of border zones – multiplex relations of power, religion and economy in South-eastern Europe, 1250-1453 CE, in: Proceedings of the 39th Annual Conference of Computer Applications and Quantitative Methods in Archaeology, "Revive the Past" (CAA) in Beijing, China (Amsterdam 2012) 381-393. For the general values of a multiplex network approach see M. Szell – R. Lambiotte – St. Thurner, Multi-relational Organization of Large-scale Social Networks in an online World. *Proceedings of the National Academy of Sciences* 107 (2010) 13636-13641: "Human societies can be regarded as large numbers of locally interacting agents, connected by a broad range of social and economic relationships. (…) Each type of relation spans a social network of its own. A systemic understanding of a whole society can only be achieved by understanding these individual networks and how they influence and co-construct each other (…) A society is therefore characterized by the superposition of its constitutive socio-economic networks, all defined on the same set of nodes. This superposition is usually called multiplex, multi-relational or multivariate network."

---

[88] For a comparison between polities within the framework of complexity theory cf. J. Preiser-Kapeller, Calculating the Middle Ages? The Project "Complexities and Networks in the Medieval Mediterranean and the Near East". *Medieval Worlds* Issue 2/2015: "Empires in Decay", 100-127.

[89] On this perspective cf. also F. Boldizzoni, The Povery of Clio. Ressurecting Economic Theory. Princeton – Oxford 2011.

**Figures**

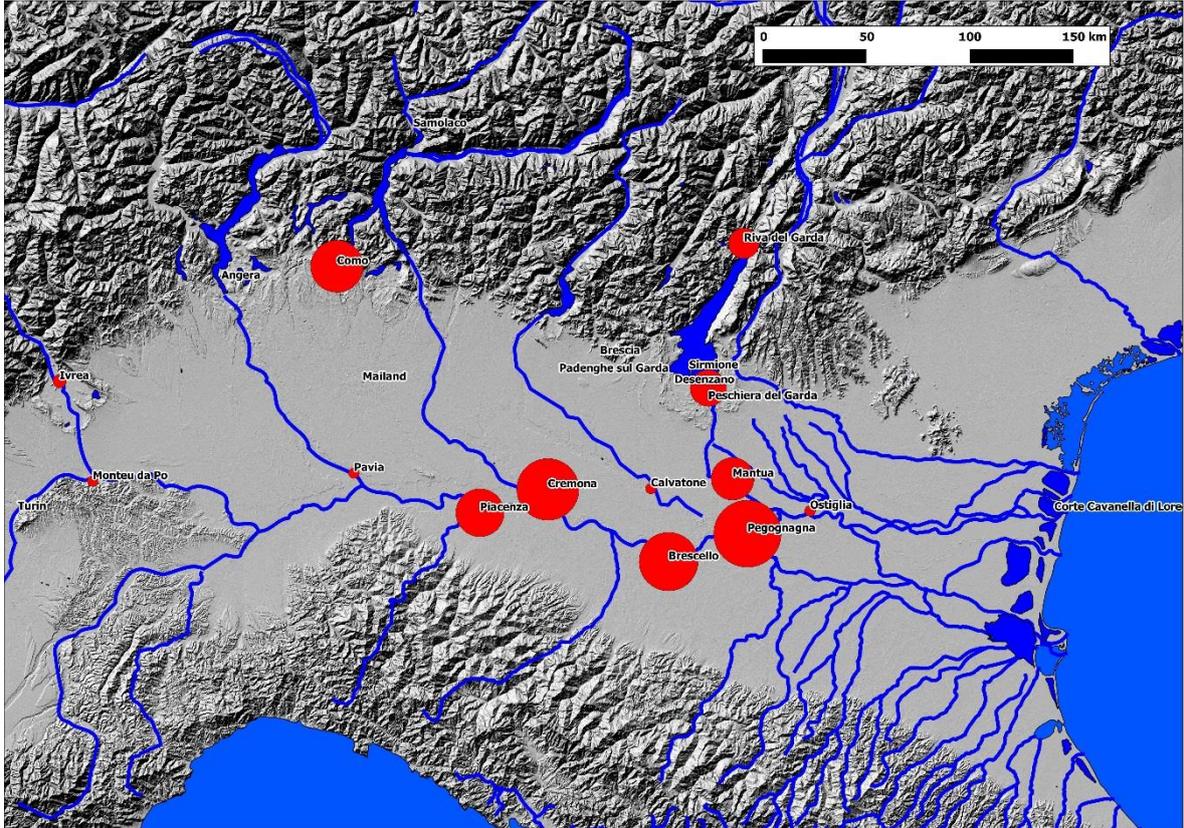

**Fig. 1:** Nodes in the network model of riverine transport in period I (1st-5th cent. CE) in the Po plain sized according to their betweenness centrality (data. L. Werther, map: J. Preiser-Kapeller, 2015).

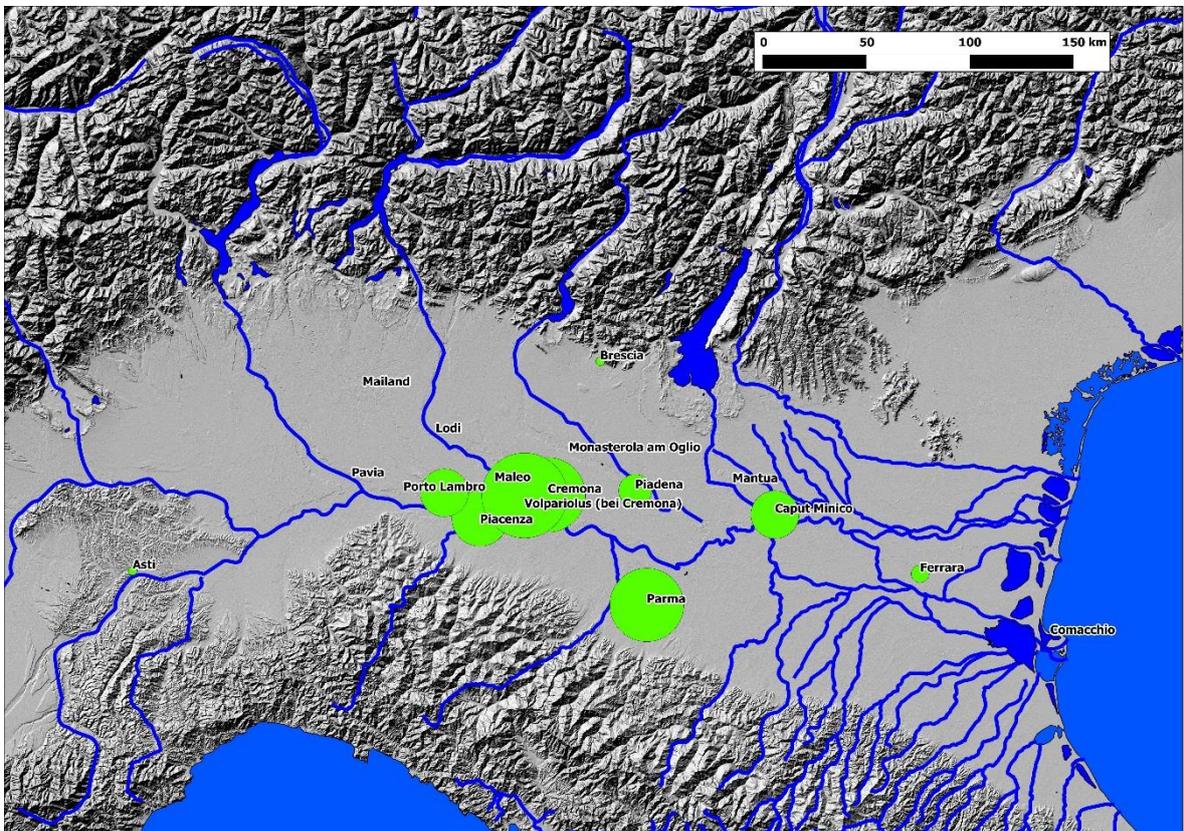

**Fig. 2:** Nodes in the network model of riverine transport in period II (6th–9th cent. CE) in the Po plain sized according to their betweenness centrality (data. L. Werther, map: J. Preiser-Kapeller, 2015).

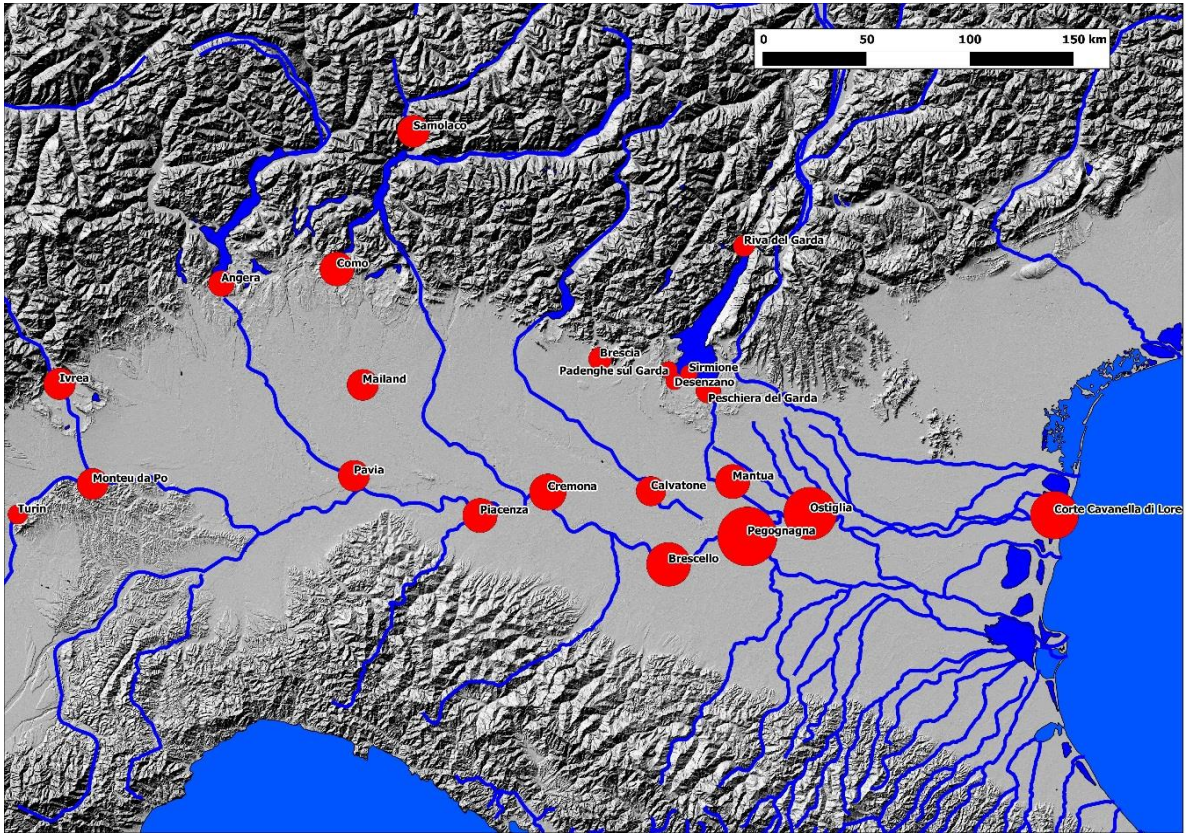

**Fig. 3:** Nodes in the network model of riverine transport in period I (1$^{st}$-5$^{th}$ cent. CE) in the Po plain sized according to their closeness centrality (data. L. Werther, map: J. Preiser-Kapeller, 2015).

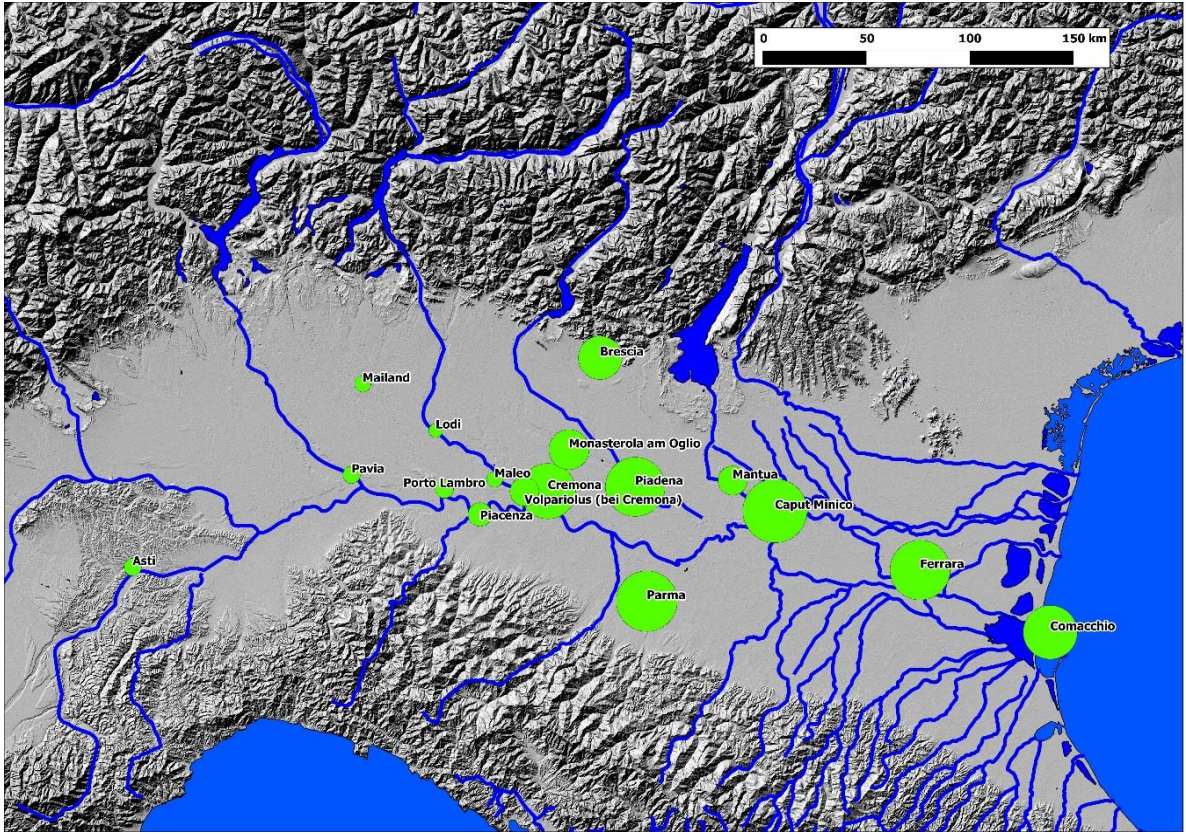

**Fig. 4:** Nodes in the network model of riverine transport in period II (6$^{th}$–9$^{th}$ cent. CE) in the Po plain sized according to their closeness centrality (data. L. Werther, map: J. Preiser-Kapeller, 2015).

|  | Po River network Period I | Po River network Period II |
|---|---|---|
| **Number of nodes** | 22 | 17 |
| **Number of edges** | 36 | 19 |
| **Density** | 0.156 | 0.136 |
| **Clustering Coefficient** | 0.505 | 0.265 |
| **Diameter** | 13 | 25 |
| **Degree Centralisation** | 0.091 | 0.070 |
| **Betweenness Centralisation** | 0.435 | 0.436 |
| **Transitivity** | 0.657 | 0.250 |
| **Circuitry (Alpha-index)** | 0.385 | 0.103 |

**Fig. 5:** Comparison of network measures for the network model of riverine transport in period I (1st-5th cent. CE) and period II (6th–9th cent. CE) (data. L. Werther, calculations: J. Preiser-Kapeller).

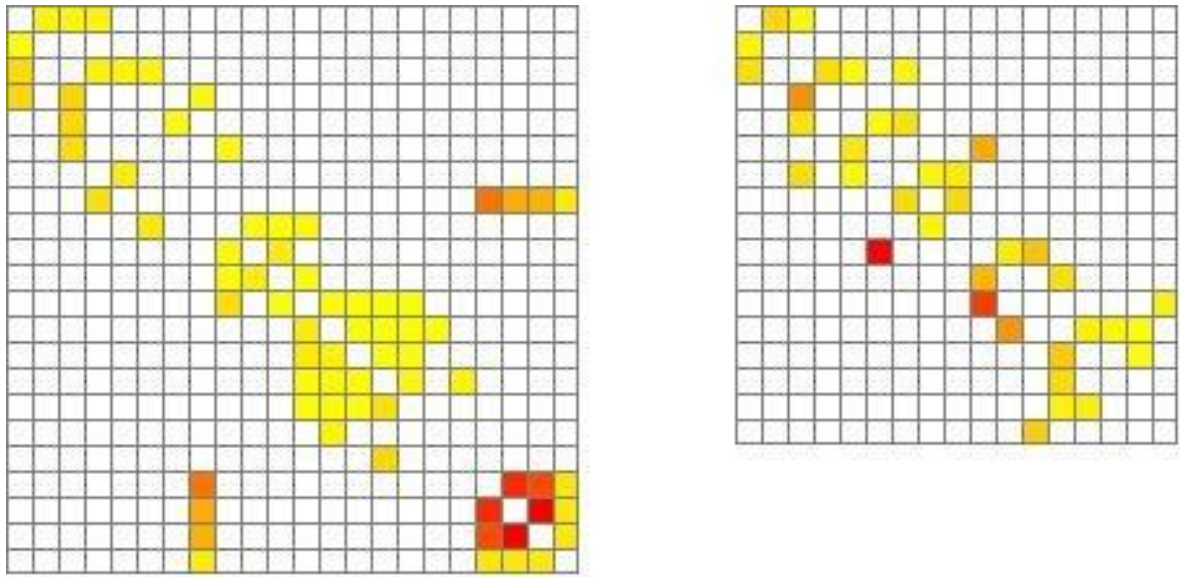

**Fig. 6:** The matrices for the network model of riverine transport in period I (1st-5th cent. CE, left) and period II (6th–9th cent. CE, right) (data. L. Werther, graphs: J. Preiser-Kapeller, 2015).

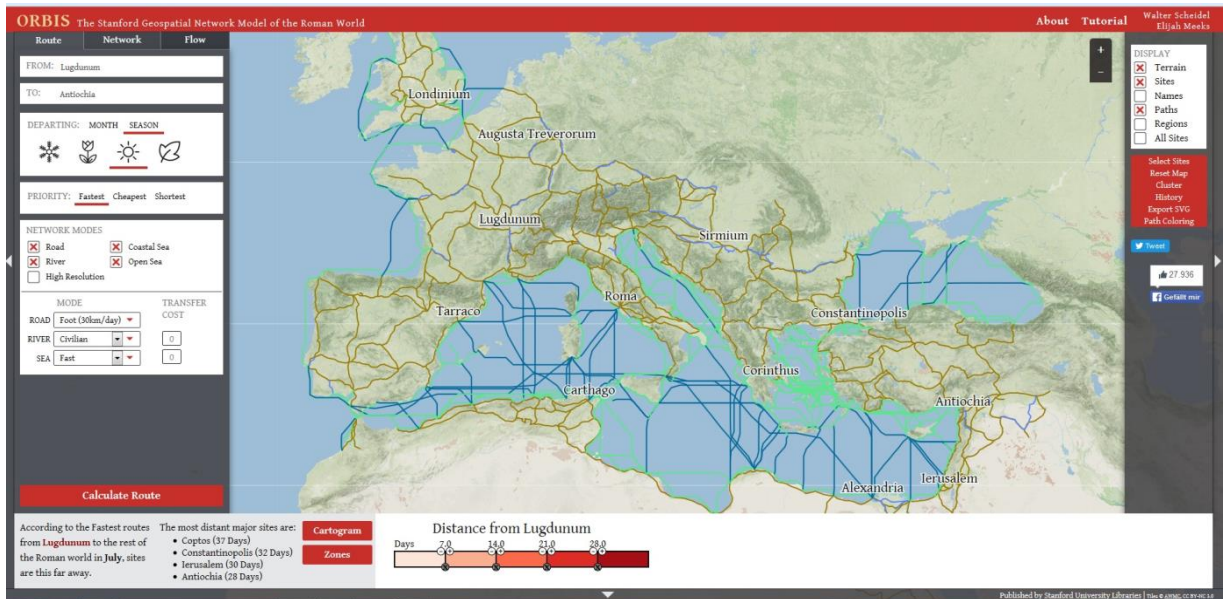

**Fig. 7**: User interface of the "ORBIS Stanford Geospatial Network Model of the Roman World" (screen shot from: http://orbis.stanford.edu/)

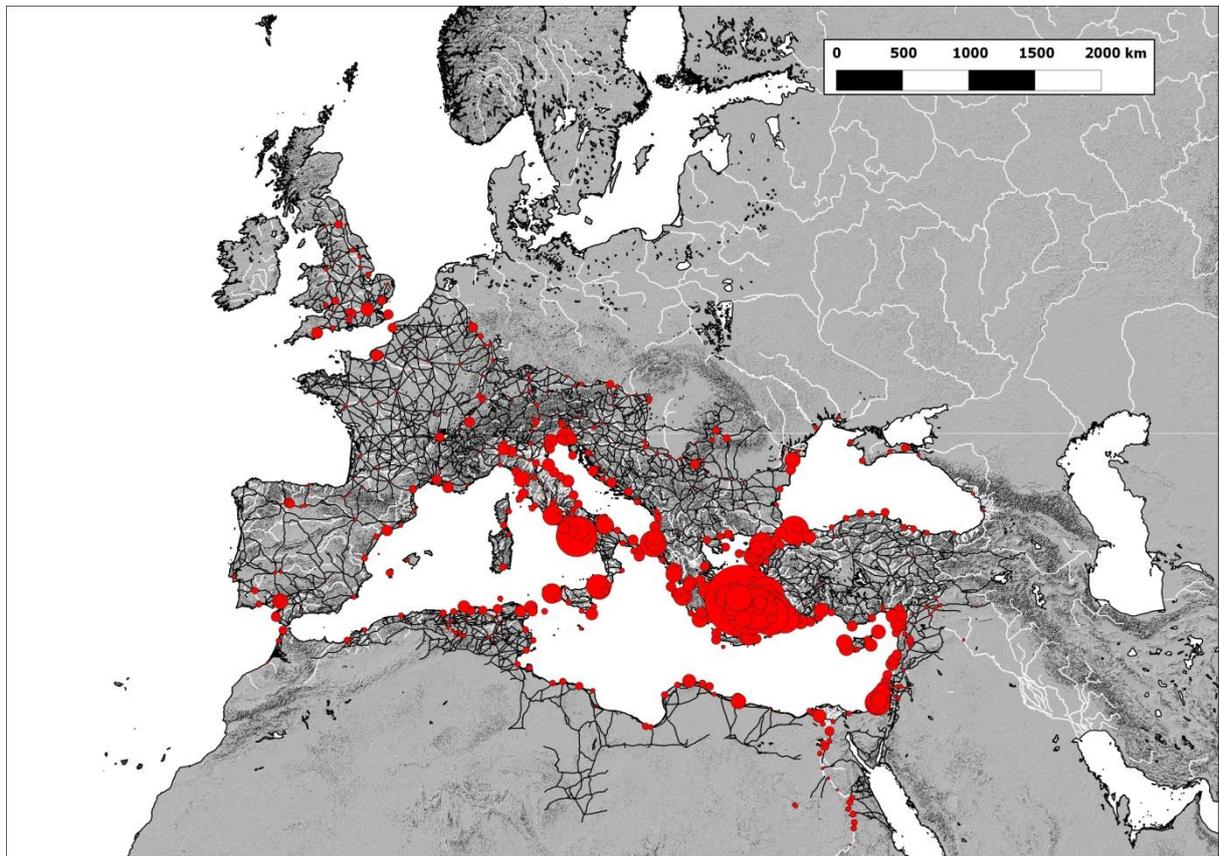

**Fig. 8:** ORBIS Stanford Geospatial Network Model of the Roman World – visualisation of the nodes (= places) sized according to their degree-centrality (analysis and map J. Preiser-Kapeller, 2015)

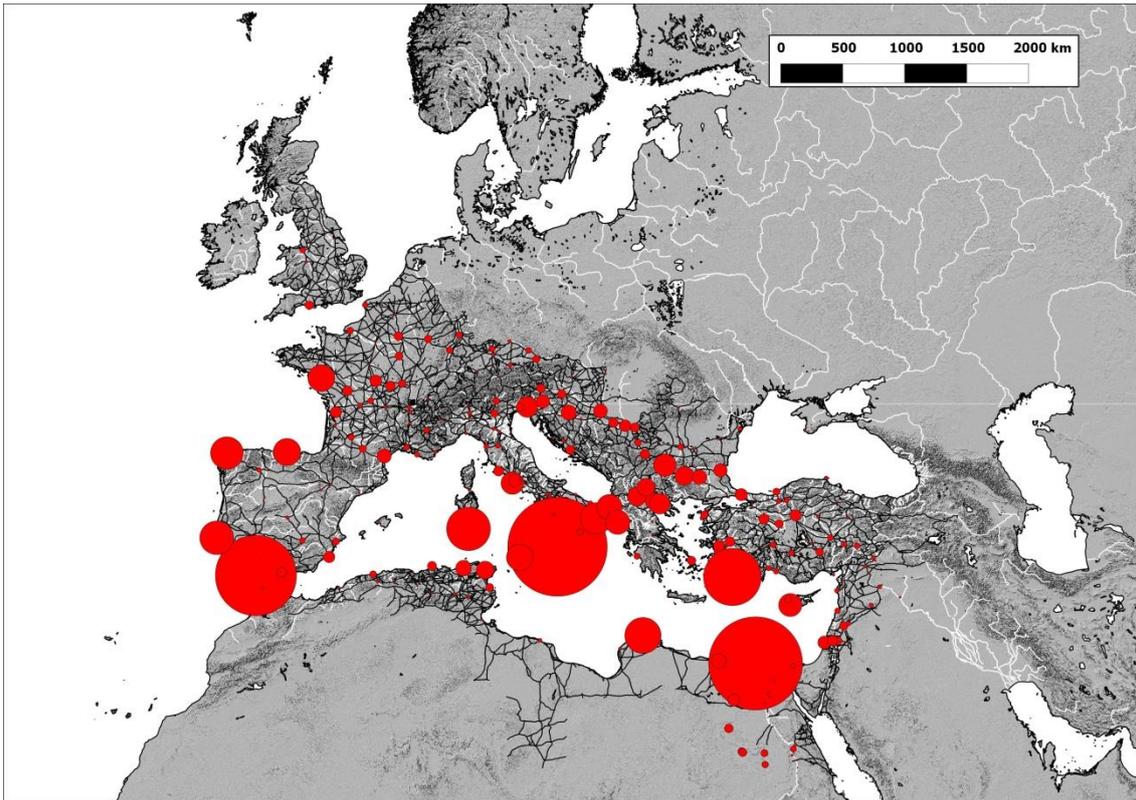

**Fig. 9:** ORBIS Stanford Geospatial Network Model of the Roman World – visualisation of the nodes (= places) sized according to their betweenness-centrality (analysis and map J. Preiser-Kapeller, 2015)

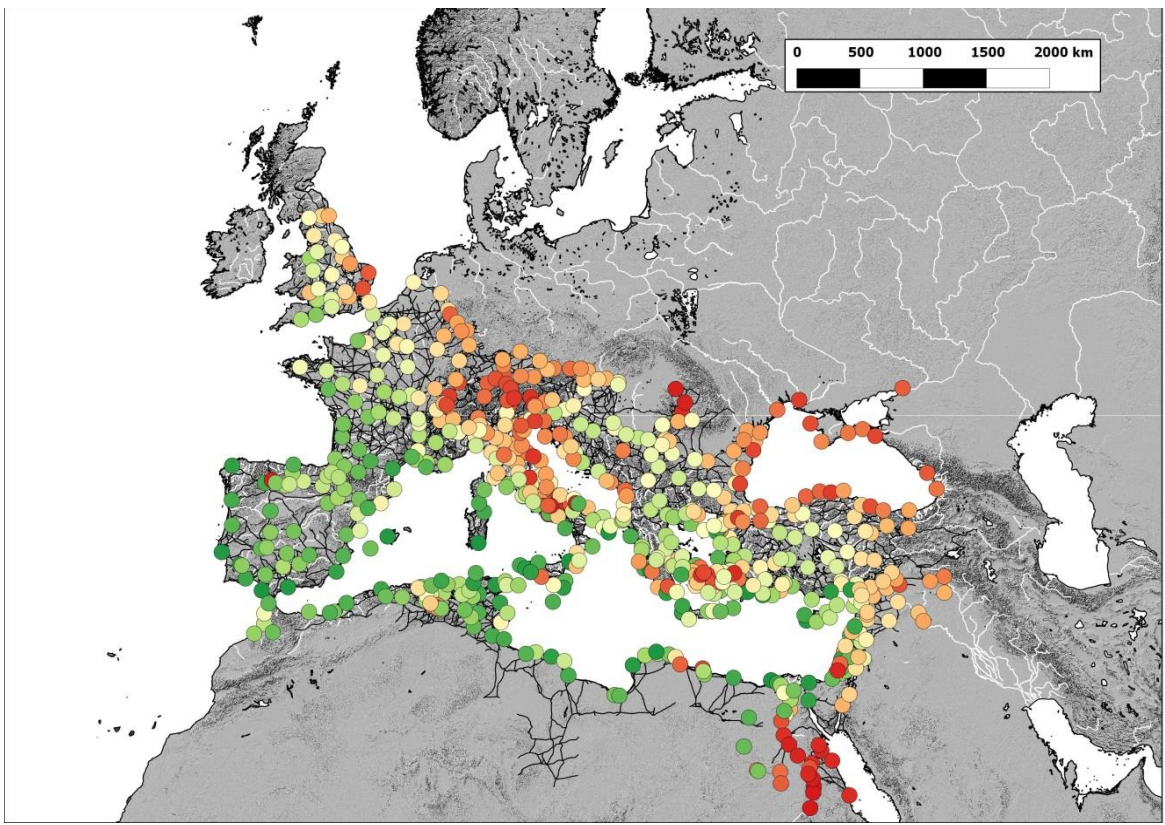

**Fig. 10:** ORBIS Stanford Geospatial Network Model of the Roman World – visualisation of the nodes (= places) coloured according to their closeness-centrality (colour scale from red/low centrality to green/high centrality; analysis and map J. Preiser-Kapeller, 2015)

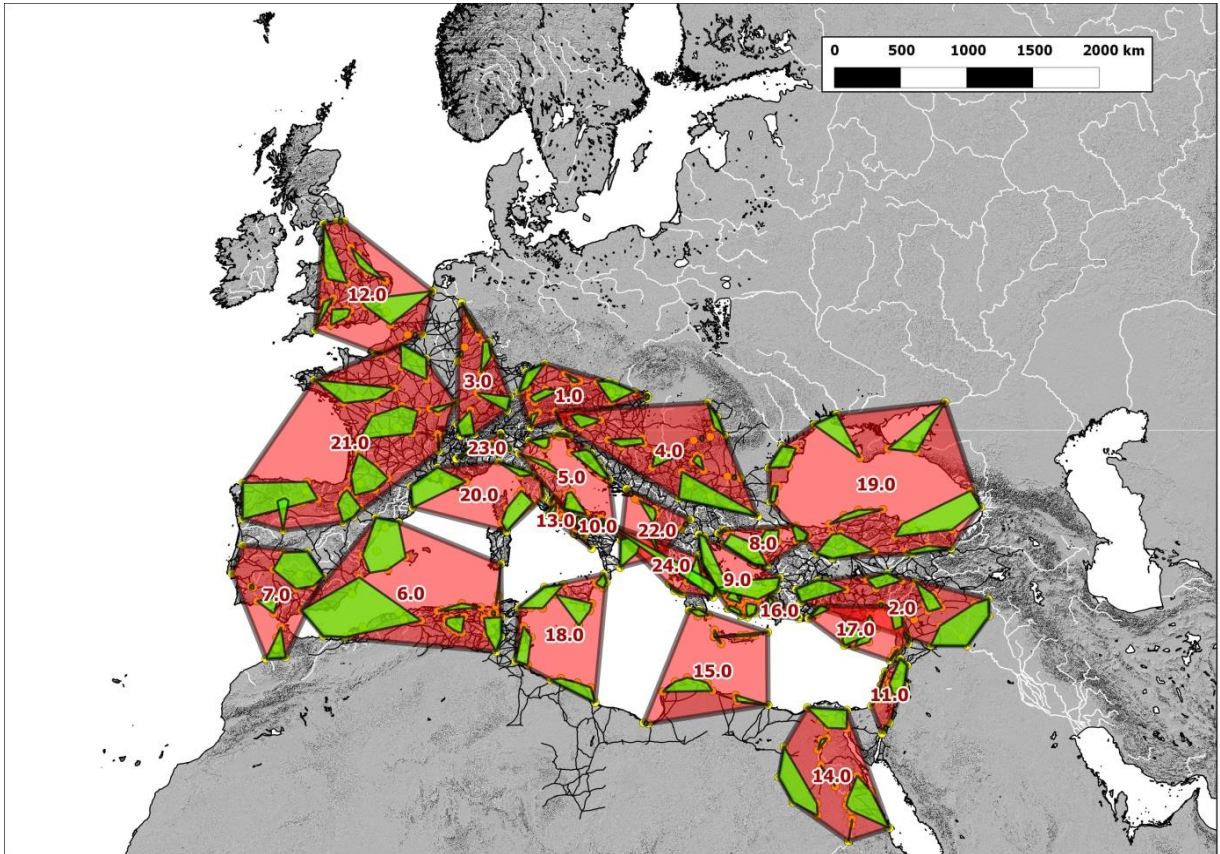

**Fig. 11:** ORBIS Stanford Geospatial Network Model of the Roman World – identification of clusters (red) and sub-clusters (green) with the help of the Newman-algorithm (analysis and map J. Preiser-Kapeller, 2015)

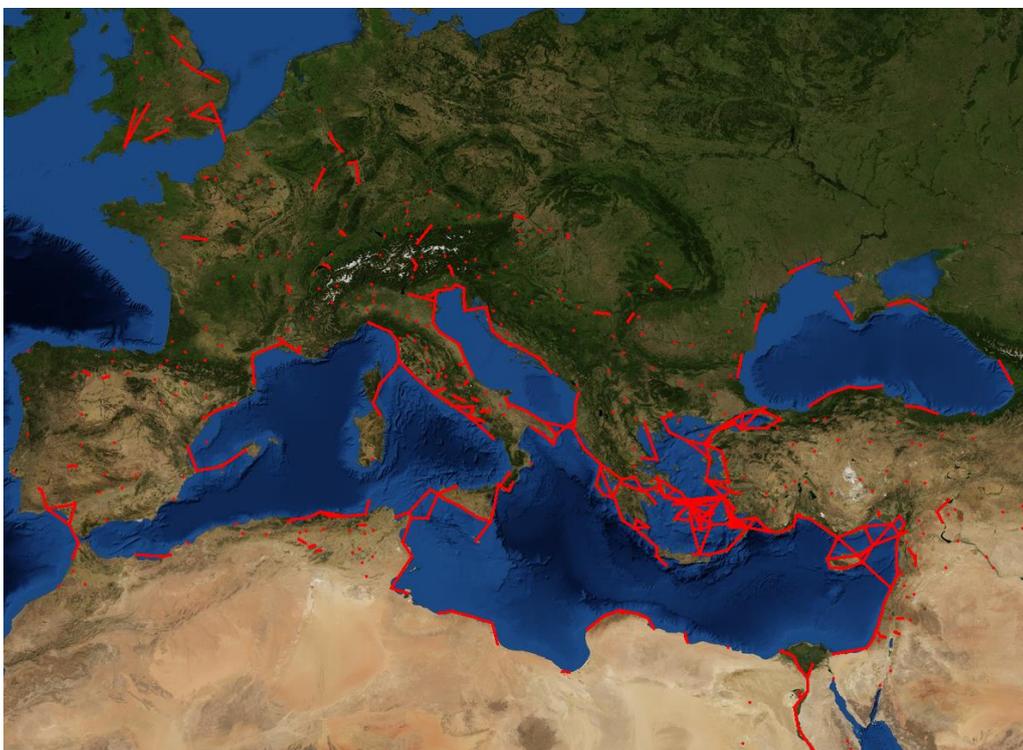

**Fig. 12:** ORBIS Stanford Geospatial Network Model of the Roman World – visualisation of routes with a "cost" of maximum one day´s journey between two places (analysis and map J. Preiser-Kapeller, 2015)

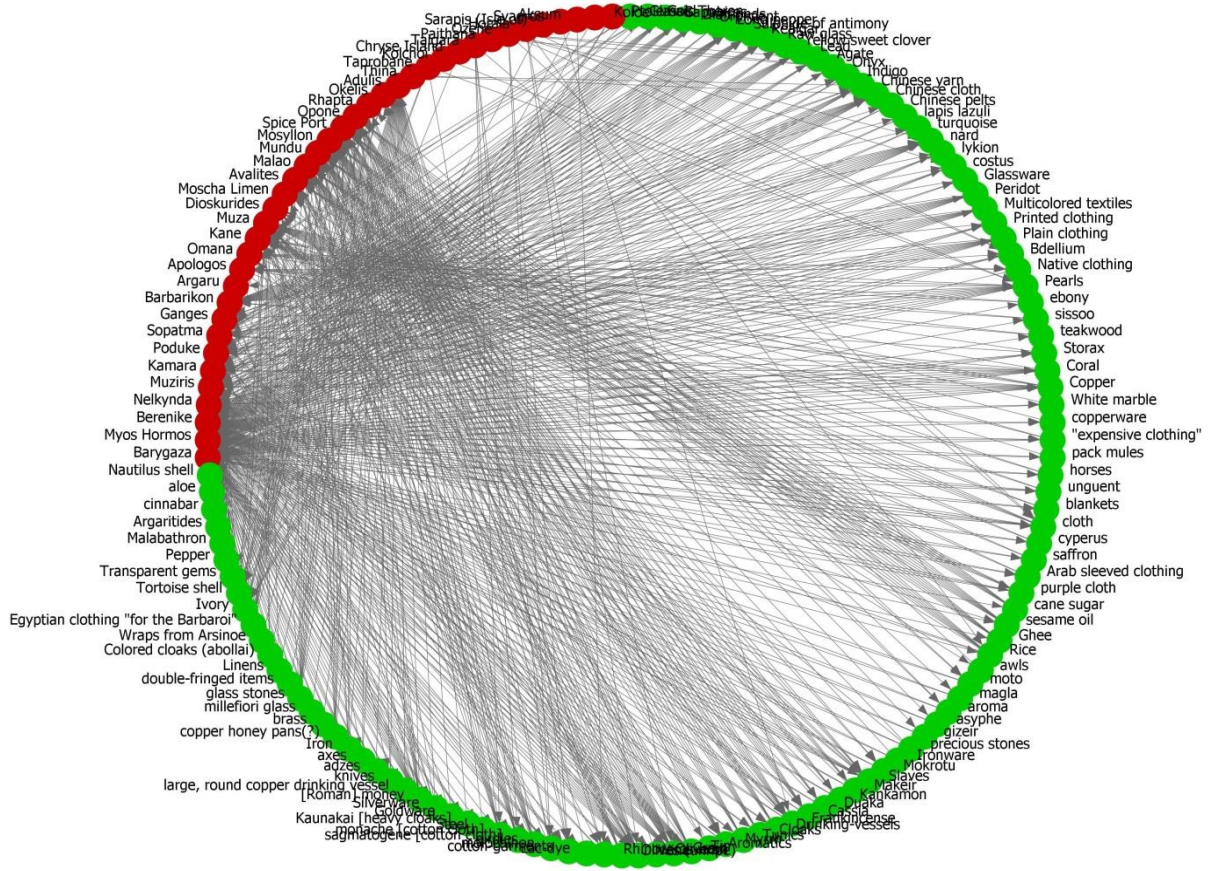

**Fig. 13:** Two-mode-network of places (red nodes) and commodities (green nodes) exported from or imported to them as narrated in the "Periplus of the Erythraean Sea" (data: E. H. Seland, http://bora.uib.no/handle/1956/11470; visualisation: J. Preiser-Kapeller, 2016)

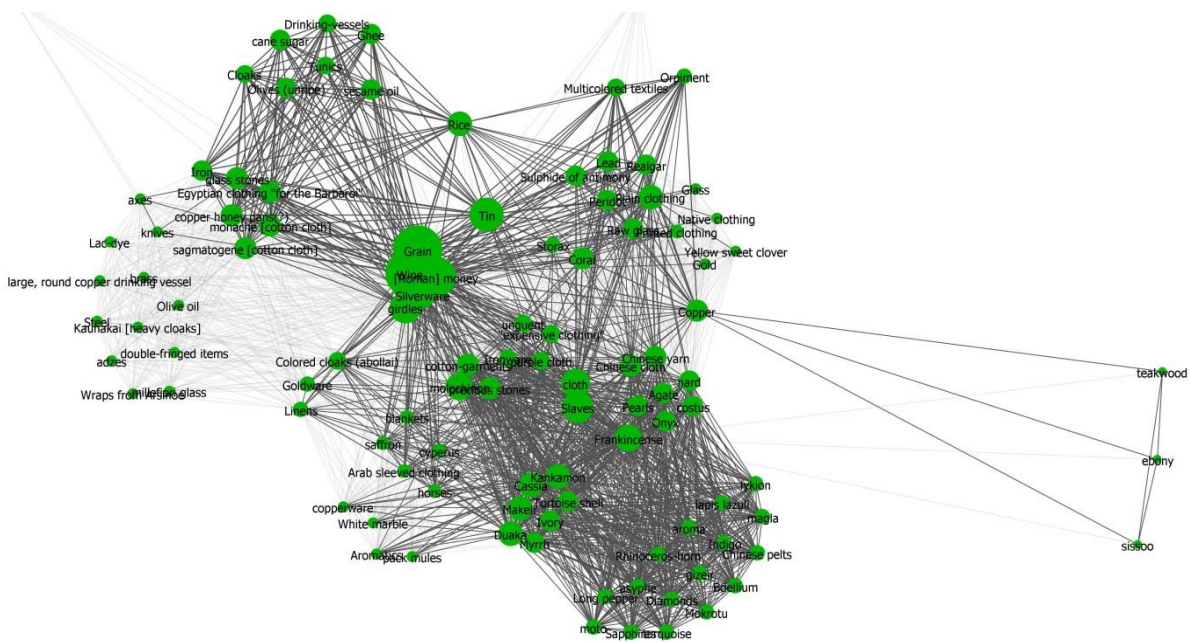

**Fig. 14:** One-mode-network of commodities due to their common export from or import to places as narrated in the "Periplus of the Erythraean Sea"; nodes sized according to their degree-centrality (data: E. H. Seland, http://bora.uib.no/handle/1956/11470; visualisation and analysis: J. Preiser-Kapeller, 2016)

**Fig. 15:** One-mode-network of commodities due to their common export from or import to places as narrated in the "Periplus of the Erythraean Sea"; nodes sized according to their betweenness-centrality (data: E. H. Seland, http://bora.uib.no/handle/1956/11470; visualisation and analysis: J. Preiser-Kapeller, 2016)

**Fig. 16:** One-mode-network of places due to their common export from or import of commodities as narrated in the "Periplus of the Erythraean Sea"; nodes sized according to their degree-centrality (data: E. H. Seland, http://bora.uib.no/handle/1956/11470; visualisation and analysis: J. Preiser-Kapeller, 2016)

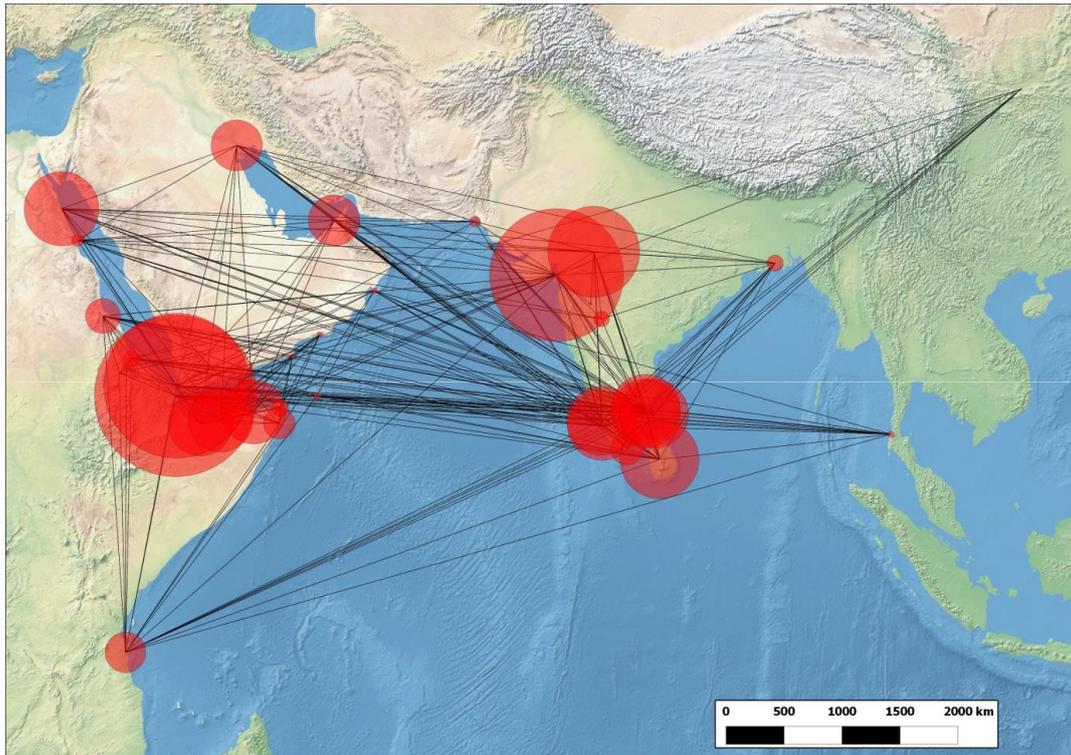

**Fig. 17:** One-mode-network of places due to their common export from or import of commodities as narrated in the "Periplus of the Erythraean Sea" visualised on a geographical map (the links indicate ties of similarity due to the exchange of the same goods, not direct ties of interaction); nodes sized according to their betweenness-centrality (data: E. H. Seland, http://bora.uib.no/handle/1956/11470; visualisation and analysis: J. Preiser-Kapeller, 2016)

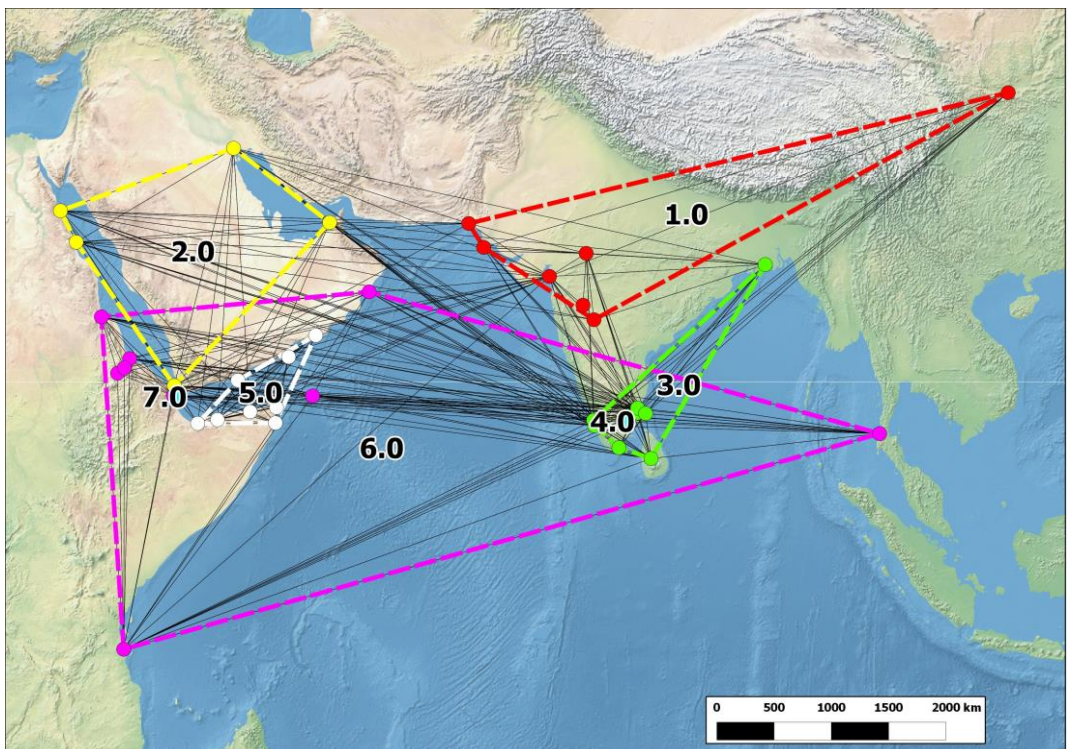

**Fig. 18:** One-mode-network of places due to their common export from or import of commodities as narrated in the "Periplus of the Erythraean Sea" visualised on a geographical map; identification of seven clusters of nodes (of different size) with the help of the Newman-algorithm (data: E. H. Seland, http://bora.uib.no/handle/1956/11470; visualisation and analysis: J. Preiser-Kapeller, 2016)

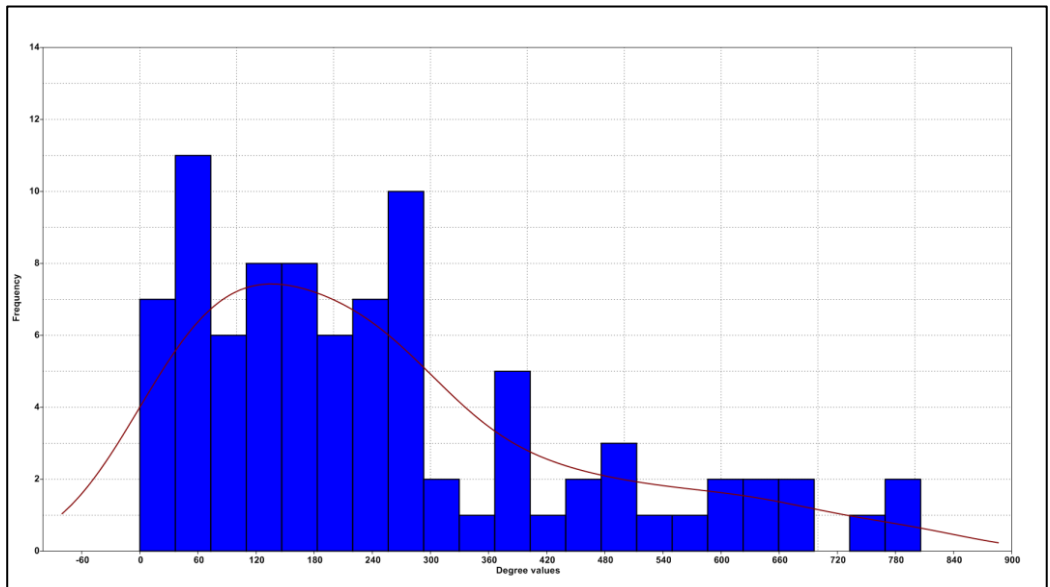

**Fig. 19:** Frequency distribution of degree values of nodes in the network model of potters from Roman potter shops (of terra sigillata) of Rheinzabern (Tabernae, ca. 150-270 CE) due to the co-occurrence of commonly used hallmarks (data: A. W. Mees, Organisationsformen römischer Töpfer-Manufakturen am Beispiel von Arezzo und Rheinzabern. 2 vol.s, Mainz 2002; graph and analysis: J. Preiser-Kapeller, 2015)

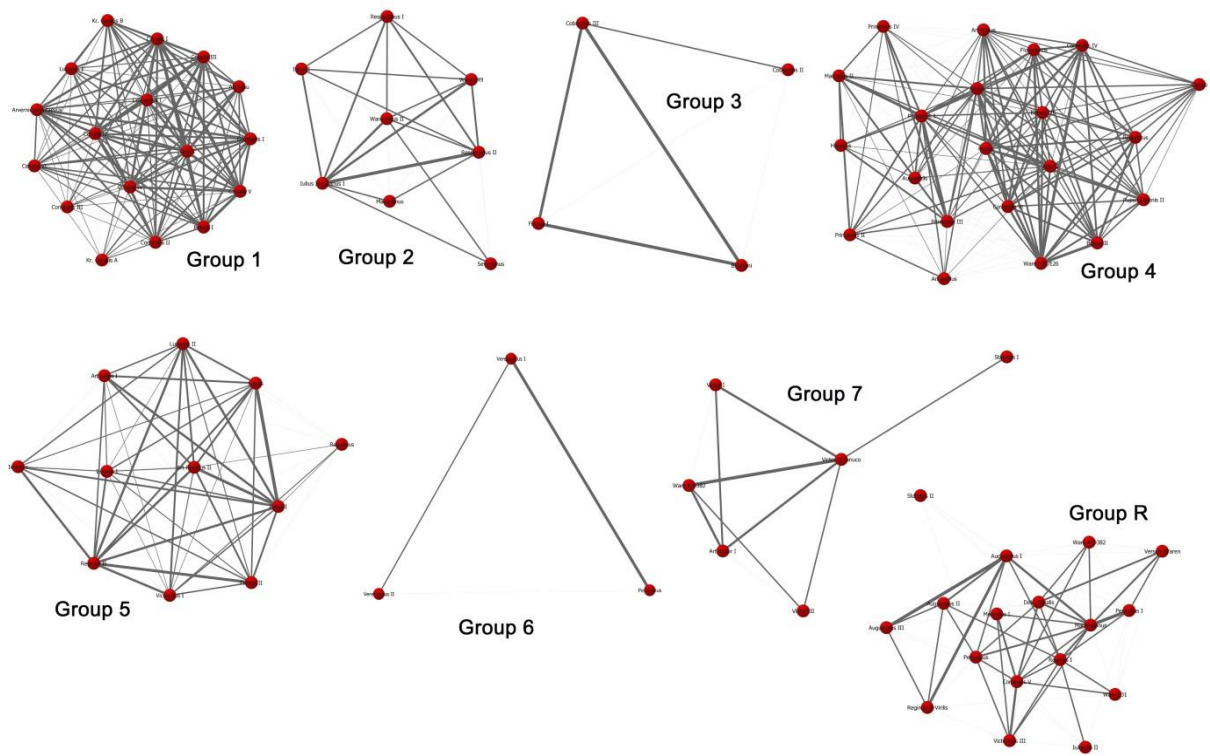

**Fig. 20:** The network model of potters from Roman potter shops (of terra sigillata) of Rheinzabern (Tabernae, ca. 150-270 CE) due to the co-occurrence of commonly used hallmarks; nodes are arranged in the eight groups of potters identified by Mees (data: A. W. Mees, Organisationsformen römischer Töpfer-Manufakturen am Beispiel von Arezzo und Rheinzabern. 2 vol.s, Mainz 2002; network modelling and graph: J. Preiser-Kapeller, 2015)

|  | Group 1 | Group 2 | Group 3 | Group 4 | Group 5 | Group 6 | Group 7 | Group R |
| --- | --- | --- | --- | --- | --- | --- | --- | --- |
| Number of nodes | 17 | 8 | 4 | 21 | 11 | 3 | 6 | 17 |
| Number of links | 135 | 25 | 6 | 185 | 54 | 3 | 10 | 70 |
| Weighted Link Sum | 2950 | 122 | 150 | 1339 | 374 | 8 | 28 | 171 |
| Density | 0.993 | 0.893 | 1 | 0.881 | 0.982 | 1 | 0.667 | 0.515 |
| Weighted Density | 0.381 | 0.189 | 0.625 | 0.193 | 0.34 | 0.667 | 0.373 | 0.126 |
| Degree Centralisation | 0.243 | 0.385 | 0.35 | 0.18 | 0.263 | 0.625 | 0.46 | 0.176 |
| Betweenness Centralisation | 0.527 | 0.482 | 1 | 0.16 | 0.622 | 0.5 | 0.4 | 0.178 |

**Fig. 21:** The network model of potters from Roman potter shops (of terra sigillata) of Rheinzabern (Tabernae, ca. 150-270 CE) due to the co-occurrence of commonly used hallmarks; structural quantitative measure for the network models of the eight groups of potters identified by Mees (data: A. W. Mees, Organisationsformen römischer Töpfer-Manufakturen am Beispiel von Arezzo und Rheinzabern. 2 vol.s, Mainz 2002; network modelling and analysis: J. Preiser-Kapeller, 2015)